\documentclass[11pt,a4paper]{article}
\pdfoutput=1
\usepackage{jheppub}

\usepackage[top=4.5cm, bottom=0.cm, outer=-0.25cm, inner=4.75cm]{geometry}
\usepackage{placeins}
\usepackage{latexsym}
\usepackage[bbgreekl]{mathbbol}
\usepackage{mathtools}
\usepackage{amsmath,amssymb,amsthm,amsbsy,amsfonts,mathrsfs}
\usepackage{hyperref}
\usepackage{multirow}
\usepackage{slashed}
\usepackage{float}
\usepackage{esint}
\usepackage{epsfig}
\usepackage{lscape}
\usepackage{titlesec}
\usepackage{enumitem}   

\usepackage{graphicx}
\usepackage[utf8]{inputenc}
\usepackage{subfigure}
\usepackage{nicefrac}
\usepackage{braket}
\usepackage{rotating}
\usepackage{afterpage}
\usepackage{bbm}
\usepackage{color,xcolor}
\usepackage{cancel}
\usepackage{stackrel}
\usepackage[framemethod=tikz]{mdframed} 
\usepackage{empheq}
\usepackage[many]{tcolorbox}
\usepackage{simplewick}
\usepackage{pgf}
\usepackage{tikz}
\usetikzlibrary{shapes.geometric}
\usetikzlibrary{decorations.markings}
\usetikzlibrary{positioning}
\usetikzlibrary{fit}
\usepackage{tkz-berge}
\usepackage{rubikcube}
\usetikzlibrary{patterns}
\usepackage{tikz-3dplot}
\usepackage[all,cmtip]{xy}
\usetikzlibrary{snakes}
\newcommand*{\circled}[1]{\lower.7ex\hbox{\tikz\draw (0pt, 0pt)
    circle (.5em) node {\makebox[1em][c]{\small #1}};}}

\tikzstyle{vertex}=[circle, draw, inner sep=0pt, minimum size=6pt]

\tikzstyle{checkpauli}=[rectangle,draw, minimum size=6pt]

\tikzstyle{diamondnode}=[diamond,draw, minimum size=6pt]

\makeatletter

\pgfkeys{/pgf/.cd,
    ellipse ratio/.code={\pgfkeyssetvalue{/pgf/ellipse ratio}{#1}},
    ellipse ratio/.initial=1
}

\pgfdeclareshape{newellipse}
{
  \inheritsavedanchors[from=ellipse]
  \inheritanchorborder[from=ellipse]
  \savedanchor\radius{%
    \pgf@y=.5\ht\pgfnodeparttextbox%
    \advance\pgf@y by.5\dp\pgfnodeparttextbox%
    \pgfmathsetlength\pgf@yb{\pgfkeysvalueof{/pgf/inner ysep}}%
    \advance\pgf@y by\pgf@yb%
    \pgf@x=.5\wd\pgfnodeparttextbox%
    \pgfmathsetlength\pgf@xb{\pgfkeysvalueof{/pgf/inner xsep}}%
    \advance\pgf@x by\pgf@xb%
    \pgfkeysgetvalue{/pgf/ellipse ratio}{\ratioscale}
    \pgfmathsetmacro\widthfactor{sqrt(\ratioscale^2+1)/\ratioscale}
    \pgfmathsetmacro\heightfactor{sqrt(\ratioscale^2+1)}
    \pgf@x=\widthfactor\pgf@x%
    \pgf@y=\heightfactor\pgf@y%
    \pgfmathsetlength\pgf@yc{\pgfkeysvalueof{/pgf/minimum height}}%
    \ifdim\pgf@y<.5\pgf@yc%
      \pgf@y=.5\pgf@yc%
    \fi%
    \pgfmathsetlength\pgf@xc{\pgfkeysvalueof{/pgf/minimum width}}%
    \ifdim\pgf@x<.5\pgf@xc%
      \pgf@x=.5\pgf@xc%
    \fi%
    \pgfmathsetlength{\pgf@xb}{\pgfkeysvalueof{/pgf/outer xsep}}%
    \pgfmathsetlength{\pgf@yb}{\pgfkeysvalueof{/pgf/outer ysep}}%
    \advance\pgf@x by\pgf@xb%
    \advance\pgf@y by\pgf@yb%
  }

  \inheritanchor[from=ellipse]{center}
  \inheritanchor[from=ellipse]{mid}
  \inheritanchor[from=ellipse]{base}
  \inheritanchor[from=ellipse]{north}
  \inheritanchor[from=ellipse]{south}
  \inheritanchor[from=ellipse]{west}
  \inheritanchor[from=ellipse]{mid west}
  \inheritanchor[from=ellipse]{base west}
  \inheritanchor[from=ellipse]{north west}
  \inheritanchor[from=ellipse]{south west}
  \inheritanchor[from=ellipse]{east}
  \inheritanchor[from=ellipse]{mid east}
  \inheritanchor[from=ellipse]{base east}
  \inheritanchor[from=ellipse]{north east}
  \inheritanchor[from=ellipse]{south east}

  \inheritbackgroundpath[from=ellipse]
}
\makeatother

\newmdenv[%
middlelinecolor=blue!20!,
middlelinewidth=1pt,
backgroundcolor=blue!10!,
roundcorner=10pt
]{identity}

\newmdenv[%
middlelinecolor=red!20!white,
middlelinewidth=1pt,
backgroundcolor=red!10!white,
roundcorner=10pt,
subtitlebelowline=true,
frametitle={Calculation},
frametitlefont={\normalfont\bfseries\sffamily\color{red!40!white}},
]{calculation}

\include{macros}

\def\bbx#1\ebx{\begin{empheq}[box={\tcbhighmath[colframe=blue!20!white,colback=blue!10!white]}]{align} #1 \end{empheq}}

\title{Real time holography for higher spin theories}

\author[]{Zezhuang Hao}

\affiliation[]{School of Mathematical Sciences, Highfield, University of Southampton, SO17 1BJ Southampton, UK}

\emailAdd{Z.Hao@soton.ac.uk}

\abstract{Real time holography is studied in the context of the embedding space formalism. In the first part of this paper, we present matching conditions for on-shell integer spin fields when going from Euclidean to Lorentzian signature on AdS background. Using the BTZ black hole as an example, we discuss various ways of lifting the physical solution from the AdS surface to the whole embedding space. The BTZ propagator for higher spin field is expressed elegantly in terms of the embedding coordinates. In the second part of the paper, we develop the proposed duality between higher spin theory and vector models. We obtain a specific map between the field configurations of these two theories in real time, so called Lorentzian AdS/CFT map. We conclude by exploring the matching conditions for higher spin fields satisfying the proposed bulk quadratic action. The physical and ghost modes can be treated independently during the Wick rotation; only physical modes are considered to be external modes.}

\begin{document}  
	\maketitle

\section{Introduction}
AdS/CFT correspondence \cite{Maldacena:1997re,gubser1998gauge,witten1998anti,Aharony:1999ti} has been brought up over two decades and most of the checks  are carried out in the large 't Hooft limit. At such limit  the bulk theory tends to stay in the low energy region thus they are described by the well studied model, for example, semiclassical field theory or supergravity. When the 't Hooft constant goes to the small limit the bulk theory will enter the high energy region thus the higher spin excitations and gravitational corrections should not be ignored while it was proposed that the spectra of string theory could be approximately described by the higher spin field theory and the boundary theory will be free, leading to the proposal that Vasiliev's higher spin \cite{Vasiliev:1990en,Vasiliev:1999ba,Vasiliev:2003ev} theory in the AdS bulk is dual to the free $O(N)$ vector model on the boundary \cite{Sezgin:2002rt,2002,Sezgin:2003pt}. The Higher spin/ Vector model duality has been verified by studying the correlation function specifically \cite{giombi2010higher,Giombi:2012ms} and recently a derivation of the duality is also given by constructing a map between the boundary fields in the bi-local form and higher spin fields on the bulk, so called AdS/CFT map \cite{deMelloKoch:2018ivk,Aharony:2020omh,Aharony:2021ovo,Aharony:2022feg}. In their work, the AdS/CFT map is constructed between the Euclidean AdS and CFT while here in this article we will first obtain the Lorentzian version of AdS/CFT map based on the study of real time holography then further study its various implications on the understanding of holography principle.\\ \\
In the first part, we developed the approach to real time holography \cite{Skenderis:2008dg,Skenderis:2008dh} from scalar case to the higher spin case by checking the match of higher spin bulk boundary propagators at the joint surface between the Lorentzian and Euclidean AdS space then obtain the match conditions for the higher spin fields. In fact, the idea of gluing the Euclidean gravitational action to the Lorentzian part has been discussed dated back to the study of wavefunction of the universe \cite{hartle1983wave} in which the Euclidean part is used to prepare the initial state of the universe so called Hartle-Hawking state. After that, the complex version of the metric has also been studied for long \cite{Gibbons:1976ue,louko1997complex,kontsevich2021wick,witten2021note} while till now the physical meaning of Euclidean saddle and its connection to the Lorentzian part are still not fully understood. However the  relation between Euclidean and Lorentzian signature in quantum field theory has been well studied and it turns out that there is a one-to-one correspondence between the Euclidean and Lorentzian QFT governed by the analytic continuation \cite{osterwalder1973axioms,osterwalder1975axioms}. Now based on the holography principle, one should also expect the existence of similar match conditions that will govern the quantum gravity theory when going between different signatures. Such conditions at low energy could be understood as the Wick rotation of the QFT on the curved background. \\ \\
Starting from the simplest condition, by making both of the Euclidean and Lorentzian action dynamical and assuming that they are smoothly connected at the joint surface, one could obtain the matching conditions for the scalar fields on the AdS background. It turns out that bulk-boundary propagator should also be joint smoothly at the surface where we perform the match. While for the higher spin theory, we choose to follow the opposite approach since the higher spin action is absent and it is easier to study the bulk-boundary propagator for higher spin particles in the embedding space \cite{Costa:2011mg,Costa:2014kfa}, which means we will obtain the match conditions by studying the bulk-boundary propagator. In section \ref{onshell matching}, we will first study the real time holography in embedding space then further present the matching condition for the higher spin fields. Moreover, if the fields are expanded in terms of the conformal basis, then we will see that the match between Euclidean and Lorentzian fields can be written as the match of the coefficients.\\ \\
More precisely, the matching condition are directly imposed on the AdS spacetime since in our case the physical space is AdS even though it is easier to the study the higher spin fields especially the higher spin bulk boundary propagator in embedding space treating the AdS as the hyperboloid embedded in a one dimension higher flat spacetime. By considering the field configuration, we can see that the physical fields on the AdS hyperboloid is obtained by the pushing back of the fields in embedding space thus there is a redundancy for the fields at the direction which is norm to the hyperboloid. In terms of coefficients we can see that there also exists extra degrees of freedom to choose the coefficients in embedding space which will give us the same higher spin fields, so called the pure gauge of coefficients. By checking the match of higher spin propagator explicitly, it turns that the pure gauge should also be fixed during the match therefore such redundancy will be eliminated and the fields will be uniquely fixed by the matching the physical fields in the chosen gauge. Moreover, we find that for the matching condition for the higher spin field, one needs to distinguish the spatial and time direction because the extra factor $i$ should be introduced along the time direction.   \\ \\
Before going into the study of Lorentzian AdS/CFT map, we analysis the difference between AdS space and embedding space. In this article, we insist that the AdS spacetime is physical and the onshell fields are solutions of corresponding equation of motion on the AdS background. For scalar fields, the equation of motion is KG equation. For the higher spin fields, they are governed by the nonlinear Vasilive's equation while here all we need to known is that the equation of motion could be reduced to the second order at linear level \cite{Bouatta:2004kk,Costa:2014kfa} similar to the KG equation. Therefore, due to the absence of dynamics for the embedding space, there are various ways for us to extend the onshell fields from the AdS hyperboloid to the embedding space. Here we will use the BTZ solution as an example to illustrate two extension methods. One is the direct extension of the field according to the AdS radius R while the other also requires the modification of the scale dimension $\Delta\rightarrow \Delta_R$. Both methods will give us smooth functions in embedding space while 
 the second approach will automatically introduce the dynamical structure to the embedding space, which could possibly make it physical. \\\\
In the second part, we establish the general principle for Lorentzian continuation for field theories then generalise the Euclidean AdS/CFT map to the Lorentzian version and then apply such map to the vector models. In the work \cite{Aharony:2020omh,Aharony:2021ovo}, the higher spin fields in the bulk are expanded by the bulk-boundary propagator as the basis while the bi-local field on the boundary are expanded by the three point correlation functions chosen as the conformal basis. The completeness of the conformal basis are provided by summing over the principal series and the map between field configuration is then manifested by establishing the map between the coefficients. After analysing the real time holography for higher spin fields on the bulk side, in order to obtain the Lorentzian AdS/CFT map, one also needs to figure out the Lorentzian conformal basis and the corresponding completeness relations so that one can decompose the boundary fields into coefficients. The Lorentzian version of the completeness relation is obtained following the principle developed in section \ref{continuation}. Moreover, by comparing the degrees of freedom from both sides, one can conclude that the AdS/CFT map is invertible at large N limit.\\ \\
At the end, we study the match conditions for the quadratic action which is treated as the effective action for higher spin fields. By mapping the action for the vector model on the boundary to the AdS bulk, it is proposed that the higher spin field theory is governed by the quadratic action provided the existence of proper gauge fixing procedure and the corresponding ghost modes contribute the unphysical modes coming from the higher order term. In section \ref{offshell}, we will just consider the scalar components of the quadratic action and present the match conditions for the quadratic action and it turns out that physical modes and ghost modes behave independently during the Wick rotation of the bulk theory.
\section{Real-Time Holography in Embedding Space}\label{onshell matching}
In this section, we will discuss real-time holographic fields within the embedding space formalism. Beginning with a brief summary of the approach to real-time holography developed in \cite{Skenderis:2008dg,Skenderis:2008dh}, which illustrates the discussions with the case of a free scalar field, we then lift the results to the embedding space and present the matching condition for higher spin fields. 
\subsection{Real-Time Holography: Review}
 In holography, when studying the field configuration or constructing the dictionary between two theories, people often choose to first specify the behaviour of the field at the boundary of AdS and then use them to write down the bulk-boundary propagator or the source of the operator which belongs to the boundary CFT \cite{witten1998anti,Freedman:1998tz,Costa:2014kfa}, for further development one can see the application of Fefferman-Graham expansion \cite{AST_1985__S131__95_0,graham1999volume,skenderis2002lecture}.\\ \\ This is enough for us to deal with Euclidean AdS/CFT since in Euclidean signature, spatial and time directions are indistinguishable and the data on the boundary will in principle encode the whole information of the field. As for the Lorentzian signature, specifying the behaviour of the field at the spatial boundary will no longer enable us to uniquely determine the bulk field due to the lack of information about the field at the far past or the far future, i.e., the boundary of the time direction. Moreover it is interesting to note that, if there is a black hole in the bulk, one can show that it is possible to determine such information concerning the initial and final states by specifying the behavior of the modes at the horizon \cite{son2002minkowski,herzog2003schwinger} .\\ \\
In fact, the problem of specifying the initial data for a gravitational theory has been discussed by Hartle and Hawking \cite{hartle1983wave} when studying the quantum gravity wavefunction of the universe. In order to specify the initial condition of the Lorentzian evolution they choose to glue part of Euclidean path integral to the initial codimension one surface of the Lorentzian spacetime. If we denote the action of the Euclidean and Lorentzian spacetime as $S_E$ and $S_L$, the weight of the quantum gravity path integral can be represented as 
\begin{equation}
    e^{-S_E}\;e^{iS_L},
\end{equation}
in which the term $e^{-S_E}$ can be treated as a norm factor resulting from the preparation of initial state. At quantum level, especially for the quantum field on the curved space time, this enables us to pick out a preferred vacuum and Hilbert space \cite{wald1994quantum,Witten:2021jzq}. \\ \\
At the same time, on the field theory side, one can also construct various contours in complex time plane to calculate the corresponding vacuum-vacuum, thermal or out of time order correlation function \cite{landsman1987real,Maldacena:2013xja}. The field theory along these contours are continuous, for example, one can see the study of analyticity of the Wightman functions \cite{osterwalder1973axioms,osterwalder1975axioms}. This implies that on the bulk side the Euclidean action and Lorentzian action should also match smoothly. Rather than regarding the Euclidean action as a norm factor, one should instead also treat the Euclidean action as the dynamical part, i.e., consider the total action $-S_E+iS_L$ , filling the contour with the bulk geometry, and impose the matching condition at the joint surface. \\ \\
Before presenting the matching conditions, here we first clarify the concept that what we mean by matching two theories together. Consider two physical systems that live in two regions labelled by I and II, the dynamics of two systems are governed by the action $S_I$, $S_{II}$, respectively. Moreover, we denote the intersection of these two regions as $\Sigma$. It could be a purely mathematical surface or a physically measurable junction, like the domain wall and we propose the condition that two theories joint smoothly at the surface $\Sigma$ to be
\begin{equation}
    \delta S_{I}|_{\Sigma}=\delta S_{II}|_{\Sigma},
    \label{mc}
\end{equation}
in which $\delta S|_\Sigma$ represents the boundary term of the variation. It is obvious if $S_{II}=S_{I}$ then $\Sigma$ will not exist physically. In our case, the region $\rm I$ and $\rm II$ now becomes spacetime with different signatures, the dynamics are governed by the action $S_L$, $S_E$.
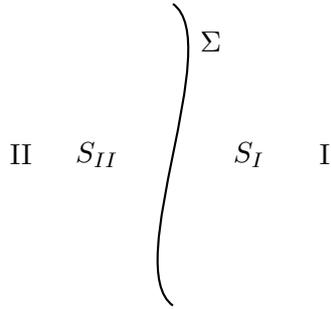
\begin{figure}
\centering
\begin{tikzpicture}
\draw (1,0) node{$S_I$};
\draw (-1,0) node{$S_{II}$};
\draw (0.5,1.5) node{$\Sigma$};
\draw (2,0) node{I};
\draw (-2,0) node{II};
\draw[thick] (0,2) .. controls (0.7,1.5)and(-0.7,-1.5) .. (0,-2);
\end{tikzpicture}
\caption{Two systems distribute in region $I$ and $II$, joint at the surface $\Sigma$.}
\label{match}
\end{figure}
As for the full quantum gravity theory, the matching condition following from \eqref{mc} is still unknown but, if we just consider the scalar field, and in the context of saddle point approximation, we have  
\begin{equation}
    \phi_E=\phi_L,\qquad i\partial_t\phi_L+\partial_\tau\phi_E=0,
\end{equation}
in which subscripts $\rm E, L$ are used to represent that the field lie along the imaginary and real part of the contour and we will stick with such convention throughout the article.\\ \\
Thus this motivates us, taking the vacuum-vacuum contour for example, to write down the duality relation \cite{Skenderis:2008dg,Skenderis:2008dh}
\begin{equation}
    \langle 0| T\;{\rm exp}\left(-i\int d^dx\sqrt{-g} \phi_{0} \mathcal{O}\right)|0\rangle=\int _{\phi\sim \phi_{0}}DgD\phi \;{\rm exp}\left(-S_E+iS_L\right),
\end{equation}
in which on the left hand side $g$ represents the bulk metric while on the right hand it represents the induced conformal structure on the boundary. As for the scalar field lives in the bulk with boundary value $\phi_0$, on the boundary CFT theory, it can be regarded as the source of the operator $\mathcal{O}$ while such correspondence are called real time gauge/gravity duality. In practice, we evaluate the right hand side at the AdS saddle point and the formula has been used to calculate the Wightman functions and produced right results. Finite temperature correlation functions are also studied provided the matching condition for thermal contour is applied on the bulk side. Moreover, it is interesting to note that, by identifying pair of sources along the thermal contour \cite{vanRees:2009rw}, one can recover the ingoing waves and retarded correlations functions when there is a black hole in the bulk.\\ \\
As for the higher spin field, the matching condition will be quite complicated. Classically, the equation of motion  for higher spin fields will be non-linear \cite{Vasiliev:1990en,Vasiliev:1999ba,Vasiliev:2003ev} while one needs to impose more physical restrictions when dealing with the matching in the context of quantum theory. One can see the discussion of massless spin two field in \cite{louko1997complex,kontsevich2021wick,witten2021note}. Here, we will only discuss the matching condition at the classic and linear level, i.e., the equation of motion are linear equations. In this case, we can write down the matching condition for higher spin fields by the investigation of the higher spin bulk-boundary propagator in embedding space.
\subsection{Holographic Scalar Fields in Embedding Space Formalism}\label{scalar}
We start from the study of scalar fields in embedding space and one can find a brief introduction of embedding space together with solutions of the KG equation in Appendix \ref{embedding}. For the embedding scalar bulk-boundary propagators, i.e., the Green function for the particle moving on the $AdS$ background with given asymptotic limit, they are deduced to be
\begin{eqnarray}
G^E_\Delta(X_1,X_2)=\frac{1}{(-2X_1\cdot X_2+i\epsilon)^\Delta_E}+Y^E_\Delta(X_1,X_2),\\\nonumber \\
G^L_\Delta(X_1,X_2)=\frac{1}{(-2X_1\cdot X_2+i\epsilon)^\Delta_L}+Y^L_\Delta(X_1,X_2),
\end{eqnarray}
in which we use $E$ and $L$ to represent different values in the two signatures while $X_1$, $X_2$ are two points in embedding space. In our expression, propagators are separated into two terms, the first is the regularised sources with proper $i\epsilon$ prescription and the second is the contribution from normalizable modes denoted as $Y_\Delta$. By solving the Klein-Gordon equation on the surface $X^2=-1$, we obtain \footnote{In fact, it is subtle to lift a function from the AdS spacetime to the embedding  space and there are various ways to do this. In the section \ref{btzs}, we will use BTZ solution as an example to discuss this in detail. }
\begin{equation}
    Y_\Delta^L(X)=\frac{1}{(2\pi)^d}\int dK \; e^{iK^\mu X_\mu/X^+}\;\theta(-K^2)\;B_L(K)\;(X^+)^{-d/2}\;J_\Delta(|K|X^+)
\end{equation}
in which $K=(w,k^i)$ for $1\le i\le d$ is the momentum space coordinate and $J_\Delta$ is the Bessel function written in terms of the scale dimension $\Delta$. $J_\Delta$  can be regarded as orthogonal basis of the normalizable modes and the coefficients $B(K)$ are determined by the boundary condition of the propagator. We can check that these two terms behave like $z^\Delta$ and $z^{d-\Delta}$ respectively when $z\rightarrow 0$ and they are two independent solutions of the asymptotic Klein-Gordon equation.\\ \\ To obtain the Euclidean version of the normalizable modes $Y^E_\Delta$, we first do the Wick rotation on $Y^L_\Delta$, i.e., taking $X^0\rightarrow -iX^0$, which will result in blow up modes when $X^0\rightarrow \pm\infty$. To get rid of this, we use the absolute value of $|X^0K_0|$ in the exponential term, which leads to
\begin{equation}
    Y^E_\Delta(X)=\frac{1}{(2\pi)^d}\int dK \; e^{(-|X^0K_0|+iX^iK_i)/X^+}\;\theta(-K^2)\;B_E(K)\;(X^+)^{-d/2}\;J_{\Delta}(|K|X^+).
\end{equation}
Given the bulk-boundary propagator, we treat them as a set of complete basis so that in general we can expand an arbitrary scalar field $\Phi_{\Delta}(X)$ as
\begin{eqnarray}
\Phi_{\Delta}(X)=\int dP\;C_\Delta(P)\;G_\Delta(X,P),
\end{eqnarray}
in which the integral means that we are suming over the null rays and $C_{\Delta}(P)$ can be treated as the coefficients of the basis while $\Delta$ is the scale dimension of the dual boundary fields, related to the mass of $\Phi_{\Delta}(X)$. Noting that there are two sets of basis thus we have two possible ways of expansion
\begin{eqnarray}
\Phi_{\Delta}^{E}(X)=\int dP\;C^E_\Delta(P)\;G^E_\Delta(X,P),\label{de1}\\\nonumber\\
\Phi_{\Delta}^{L}(X)=\int dP\;C^L_\Delta(P)\;G^L_\Delta(X,P)\label{de2},
\end{eqnarray}
\begin{figure}
\begin{tikzpicture}
\draw[thick,->] (-3,0) --(0,0);
\draw[thick] (-3,0) --(3,0);
\draw[thick] (3,-3) --(3,0);
\draw[thick,<-] (-3,0) --(-3,1.5);
\draw[thick] (-3,1.5) --(-3,3);
\draw (0,0.3) node{o};
\draw[dashed,thick](-3,1)..controls(-2.8,0.2)..(-2,0);
\draw[thick,<-] (3,-3) --(3,-1.5);
\draw (0,-0.3) node{$t$};
\draw (3.3,-1.5) node{$\tau$};
\draw (-2.7,1.5) node{$\tau$};
\draw (3,0.5) node{$T$};
\draw (-3,-0.5) node{$-T$};
 \filldraw [red] (3,0) circle (2pt);
 \draw(0,-3.5) node{(a)};
\end{tikzpicture}
\hspace{7.0em}
\begin{tikzpicture}
\draw[thick] (-1.,-3) --(-1,3);
\draw[thick] (1,-3) --(1,3);
\draw[thick] (-1,0) --(1,2);
\draw[thick] (-1,0) --(1,-2);
\draw[thick,red] (-1,0) .. controls (0,0.8) .. (1,1);
\draw[thick,->] (1.3,-0.5) --(1.3,0.5);
\draw (1.5,0) node{$t$};
\draw (1.9,1.7) node{$t=T$};
\draw[thick,<-] (0.4,0.86) .. controls (0.8,1.3)and(1.1,0.9) .. (1.4,1.4);
\draw[blue,thick] (-1,2)--(1,2);
\draw[blue,thick] (-1,-2)--(1,-2);
\draw(0,-3.5)node{(b)};
\end{tikzpicture}
\caption{(a) A toy Wick rotation contour, vertical lines represent the imaginary time $\tau$ while horizontal lines represent the real time $t$. The red point at $(T,0)$ is the matching surface $\Sigma$. The dashed curve means that we can deform the contour apart from the axes in principle. (b) The global $AdS_2$ is illustrated by the strip while our embedding coordinate covers the region between two blue lines, which is a hyperboloid. Moreover, the Poincar$\acute{{\rm e}}$ coordinates covers half of the hyperboloid showed in the triangle. The red line represents the matching surface $\Sigma$ at $t=T$.}
\label{contour}
\end{figure}
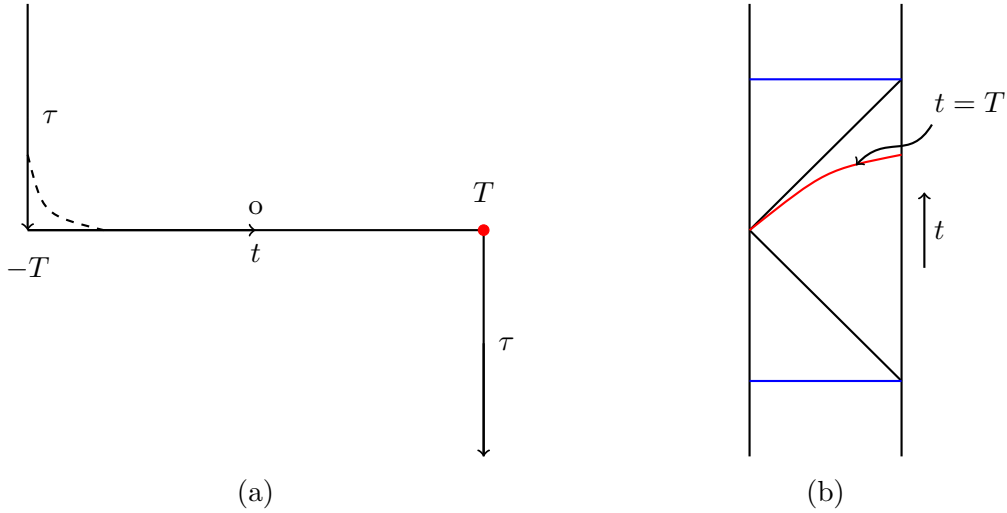
where one is for imaginary time while the other is for real time. We should keep in mind that these two kinds of expansions only work in their own proper region. Taking the contour in Fig.\ref{contour} for example, which starts from $(-T,+\infty)$ and ends at $(T,-\infty)$ with two corners at $(\pm T,0)$,  we can interpret that two vertical lines are used to prepare the initial and final quantum state and the horizontal line represents the evolution of time from $-T$ to $T$.\\ \\ To study the behavior of the fields along this contour, we should apply the Euclidean expansion $\Phi_E$ on $\tau$ axes and the Lorentzian expansion $\Phi_L$ on $t$ axes. For the contour going through the plane, like the dashed line around $-T$, these two kinds of basis will mix and it will go beyond our discussion. Since the contour with corners are not smooth, to make the fields $\Phi^E_\Delta$ and $\Phi^L_\Delta$ consistent along the whole contour, matching conditions should be imposed at the singular surface. Consider the surface at $t=T$ and we denote it as $\Sigma$ , illustrated in Fig.\ref{contour}, the matching conditions \cite{Skenderis:2008dh,Skenderis:2008dg} are
\begin{eqnarray}
\Phi^E_\Delta(X)\arrowvert_{\Sigma}=\Phi^L_\Delta(X)\arrowvert_{\Sigma}\label{m1},\\\nonumber\\
\partial_\tau\Phi^E_\Delta(X)\arrowvert_{\Sigma}+i\partial_t\Phi^L_\Delta(X)\arrowvert_{\Sigma}=0\label{m2},
\end{eqnarray}
and these two conditions enable us to solve $C^L_\Delta$, $C^E_\Delta$ and $B_L$, $B_E$. Since the basis of the source and the normalizable modes are independent, we will deal with $B$ and $C$ separately. First for the source terms, with the help of \eqref{de} and \eqref{dl}, we choose to push forward the derivative of the time to the embedding space
\begin{eqnarray}
\partial_\tau\frac{1}{(-2X\cdot P)^\Delta_E}=\frac{2\Delta}{(-2X\cdot P)^{\Delta+1}_E}(-X^0P^++X^+P^0)\label{p1},\\\nonumber \\
\partial_t\frac{1}{(-2X\cdot P)^\Delta_L}=\frac{2\Delta}{(-2X\cdot P)^{\Delta+1}_L}(X^0P^+-X^+P^0)\label{p2}.
\end{eqnarray}
Moreover, in order to use the equation \eqref{m1} and \eqref{m2} to solve the coefficients, we need to figure out the form of a function when it is restricted on the surface $\Sigma$ from the Euclidean and Lorentzian point of view. For Lorentzian signature, since the contour lies exactly along the real $t$ axes, the restricted function on $\Sigma$ means that we take $t=T$. For the Euclidean signature, since the contour shifts away from the pure imaginary axes $\tau$ and in order to reflect such shift in the theory, we need to make the variable in the function shift as $X^0\rightarrow X^0+iT$ and then set $\tau=0$. Moreover, after taking the coordinate rotation $P^E_0 =iP^L_0$ into consideration and substituting  \eqref{p1}, \eqref{p2} into the matching condition equations, we find the solution should be
\begin{equation}
\label{scalar match}
C^E_\Delta=C^L_\Delta.
\end{equation}
As for the normalizable term, by directly comparing the integrands, from \eqref{m1} we have
\begin{equation}
    B_L(K)+B^-_L(K)=B_E(K)+B_E^-(K),
\end{equation}
while from \eqref{m2} we obtain\footnote{In fact here we take $T=0$ otherwise there will be a phase factor $e^{\pm iwT}$ in front of the coefficients. Since $B$ should only depend on the boundary conditions rather than the choice of contour, it will not change the result.}
\begin{equation}
    B_L(K)-B^-_L(K)=B_E(K)+B_E^-(K),
\end{equation}
in which we define $B^-(w,k^i)=B(-w,k^i)$ and these two equations will lead to
\begin{equation}
     B_E=B_L=0,
\end{equation}
which tells us normalizable modes do not contribute to the bulk-boundary propagator.

\subsection{Matching Conditions for Higher Spin Fields}
Now, we are going to discuss the matching condition of higher spin fields and lift them to the embedding space as briefly introduced in Appendix \ref{embedding} and extensively studied in \cite{weinberg2010six,Costa:2011mg,Costa:2014kfa,Parisini:2022wkb}.  The $\rm AdS_{d+1}$ spacetime is regarded as a hyperboloid $X^2=-1$ in the embedding space, as we have introduced before. Here we use the embedding coordinates in the light cone gauge $X^A=(X^+,X^-,X^\mu)$ and the $\rm AdS$ Poincar$\acute{{\rm e}}$ coordinates $y^a=(y^\mu,z)$. Moreover, the Jacobian matrix between these two coordinates is given by  
\begin{equation}
    \frac{\partial X} {\partial y^\mu}=\frac{1}{z}(\;0,\;2y_\mu,\; \delta^{\nu}_{\mu}\;)\qquad {\rm and }\qquad \frac{\partial X} {\partial z}=(-\frac{1}{z^2},\;1,\; 0,\dots,0),
\end{equation}
which tell us the rule to push back a vector $H_A$ in embedding space to a vector $H_a$ in $AdS$ space as a submanifold thus we have \begin{equation}
    H_a(X)=\frac{\partial X^A} {\partial y^a}H_A(X).
\end{equation}
Noting that the dimension of $\rm AdS$ is lower than the embedding space, there is a redundancy when we transform form $H_A$ to $H_a$. To construct a one-to-one correspondence between symmetric tensors in these two spaces, we should impose the transverse condition $X^A H_A(X)=0$ to eliminate the extra degree of freedom while we can understand it as restricting the vector $H_A$ tangent to the $\rm AdS$ submanifold.\\ \\
Before writing down the matching conditions in embedding space, based on the study of scalar field matching, we first propose the two conditions in the $\rm AdS$ space 
\begin{eqnarray}
H_{a}^{E}(X)|_\Sigma=H_{a}^{L}(X)|_\Sigma, \\\nonumber\\
\partial_\tau H^E_a(X)|_\Sigma+i\partial_t H^L_{a}(X)|_\Sigma =0.
\end{eqnarray}
These two equations specify conditions at the joint surface $\Sigma$ up to the first order. Since perturbatively physical equations with integer spin, like the curved background Maxwell equation, are second order differential equations\footnote{For the fields with half integer spin, the equation of motion will become first order while one needs to consider the match of spin structure between two different signatures.   }; the above conditions at the boundary enable us to determine the tensor field in a unique way. As for the first matching condition, using the Jacobian matrix, we can write it in terms of the embedding coordinate as
\begin{equation}
   \left( \frac{\partial X^A}{
   \partial y^a}\right) ^E H^E_A(X)= \left(\frac{\partial X^A}{
   \partial y^a}\right)^L H^L_A(X),
    \label{hsm1}
\end{equation}
in which we omit the joint surface restriction and again we use $E, L$ to distinguish Jacobian matrices in Euclidean and Lorentzian signature, respectively.  As for the second matching condition, since the derivative with respect to time $\partial_t=\partial_\tau=\partial_0$ is involved, we need to study the matching conditions in different spacetime directions separately. Taking the Euclidean signature for example, for $a\neq 0 $, we have
\begin{equation}
    \partial_\tau\left(\;\frac{\partial X^A}{\partial y^a} H_A(X)\;\right)=\frac{\partial X^A}{\partial y^a}\;\partial_{\tau}H_A(X),\label{j1}
\end{equation}
while for $a=0$
\begin{equation}
\partial_\tau\left(\;\frac{\partial X^A}{\partial y^0} H_A(X)\;\right)=\frac{\partial X^A}{\partial y^0}\;\partial_{\tau}H_A(X)+2X^+H_+,\label{j2}
\end{equation}
which tells us that $\partial_\tau$ will commute with the Jacobian matrix except for the $y^0=\tau$ direction in which we get the extra $2X^+H_-$ contribution. This leads us to define an operator valued matrix $M$ as
\begin{equation}
M:=\frac{\partial X}{\partial y}\partial_\tau+
\left(
\begin{array}{ccc}
2X^+ & 0\\ \\
0& 0
\end{array}
\right)=\frac{\partial X}{\partial y}\partial_\tau+\mathcal{T},
\label{ma1}
\end{equation}
in which the second term $\mathcal{T}$ is a $(d+1)\times (d+2)$ matrix which has $(0,\dots,d-1,z)$ as rows and $(-,+,0,\dots,d-1)$ as columns while $(0,-)$ is the only nonzero element. With the help of matrix $M$, we can write \eqref{j1} and \eqref{j2} in a compact form as
\begin{equation}
     \partial_\tau \left(\;\frac{\partial X^A}{\partial y^a} H_A(X)\;\right)=M_a^A\; H_A,
\end{equation}
and the second matching condition can be written as
\begin{equation}
    M^E\cdot H^E(X)+iM^L\cdot H^L(X)=0,
    \label{hsm2}
\end{equation}
in which $M^E$ is the matrix  we have already discussed in \eqref{ma1} acting on the column vector $H_A$ and $M^L$ is
\begin{equation}
M^L=\left(\frac{\partial X}{\partial y}\right)^L\partial_t-
\left(
\begin{array}{ccc}
2X^+ & 0\\ \\
0& 0
\end{array}
\right).
\end{equation}
Equation \eqref{hsm1} together with \eqref{hsm2} give us matching conditions for the vector fields while it is straight forward to generalise them to the high spin fields. Before doing that, we first hide the spin index into the polynomial via the operator $K_A$ \cite{Costa:2014kfa}
\begin{equation}
    H_{A_1\dots A_J}(X)=\frac{1}{J!(\frac{d-1}{2})_J}K_{A_1}\dots K_{A_J}H(X,W),
\end{equation}
in which $H(X,W)=W^{A_1}\dots W^{A_J}H_{A_1\dots A_J}(X)$ is a polynomial in terms of $W$. Moreover, we should note that $K_A$ only involves the variable $W$ provided $H_{A_1\dots A_j}$ is a symmetric traceless tensor.  Thus $K_A$ acts on the polynomial $H(X,W)$ independently  and we can write higher spin field matching conditions in terms of the embedding polynomials as
\begin{eqnarray}
   \left(\frac{\partial X^{A_1}}{\partial y^{a_1}}K_{A_1}\right)^E\dots \left(\frac{\partial X^{A_J}}{\partial y^{a_J}}K_{A_J}\right)^E H^E(X,W)\\\nonumber\\=\left(\frac{\partial X^{A_1}}{\partial y^{a_1}}K_{A_1}\right)^L\dots \left(\frac{\partial X^{A_J}}{\partial y^{a_J}}K_{A_J}\right)^L H^L(X,W)\nonumber
\end{eqnarray}
and 
\begin{eqnarray}
   \sum _{i=1}^J \left(\frac{\partial X^{A_1}}{\partial y^{a_1}}K_{A_1}\right)^E\dots \left(T_{a_i}^{A_i}K_{A_i}\right)^E\dots\left(\frac{\partial X^{A_J}}{\partial y^{a_J}}K_{A_J}\right)^E H^E(X,W)\\\nonumber\\+i \sum _{i=1}^J \left(\frac{\partial X^{A_1}}{\partial y^{a_1}}K_{A_1}\right)^L\dots \left(T_{a_i}^{A_i}K_{A_i}\right)^L\dots\left(\frac{\partial X^{A_J}}{\partial y^{a_J}}K_{A_J}\right)^L H^L(X,W)=0,\nonumber
\end{eqnarray}
in which we take $T^{A_J}_{a_J}=M^{A_J}_{a_J}$ for short. At first sight, the matching conditions in embedding space are more complicated than those in $\rm AdS$ space since more operators and transform matrices are involved but we should note that the tensor field $H_{a_1\dots a_J}$ will be simplified a lot once we write it into the form of polynomial $H(X,W)$. 
 \subsection*{Examples}
 Here we will see how the matching conditions work based on the investigation of the fields with spin $J=1$ and $J=2$. During the discussion, we use the spin $J$ bulk-to-boundary propagator of dimension $\Delta$ written as
 \begin{equation}
     G_{\Delta,J}(X,P;W,Z)=\frac{((-2P\cdot X)(W\cdot Z)+2(W\cdot P)(Z\cdot X))^J}{(-2P\cdot X)^{\Delta+J} }\label{bbo},
 \end{equation}
 in which $X, P$ are points live on the bulk and boundary associated with the polynomial variables $W,Z$. The above expression is determined by imposing the boundary condition on bulk Green function
 \begin{equation}
    \lim_{z_2\to 0}z_2^{-\Delta}G_{\Delta,J}(X_1,X_2;Z_1,Z_2)=G_{\Delta,J}(X_1,X_2^{\infty};Z_1,Z_2)
    \label{lim}.
 \end{equation}
 Here we are abusing the notion $G_{\Delta,J}$. On the left it represents the bulk Green function while on the right it represents a bulk-boundary operator. The limit $z\to 0$ means that we are sending a bulk point $X=\frac{1}{z}(1,z^2+y^2,y^\mu)$ to the boundary $X^{\infty}=(1,y^2,y^\mu)$ thus the explicit physical meaning of $G_{\Delta,J}$ depends on the position of its second point. Moreover, after taking the limit in \eqref{lim}, we can see the formula in $\eqref{bbo}$ only encodes the information about the sources while the information about the normalizable modes proportional to $z^{d-\Delta}$ is absent.\\ \\
 As we have mentioned before, to extract the tensor from a given polynomial, we need to act the operator \cite{Costa:2011mg,Costa:2014kfa}
 \begin{equation}
     D^A_Z=\left(\frac{d}{2}-1+Z\cdot\frac{\partial}{\partial Z}\right)\frac{\partial}{\partial Z_A}-\frac{1}{2}Z^A\frac{\partial^2}{\partial Z\cdot \partial Z}
 \end{equation}
 and 
 \begin{equation}
     K^W_A=\left( \frac{d-1}{2}+W\cdot\frac{\partial}{\partial W}\right)\frac{\partial}{\partial W^A}
 \end{equation}
 on the polynomial corresponding to the bulk and boundary points, respectively. In the following two examples, we will try to deal with the field of spin $J=1$ and $J=2$ separately and the above operators will be used extensively together with the higher spin field expansion
 \begin{equation}
    H(X,W)=\int \frac{dP}{J!(\frac{d}{2}-1)_J}C_{\Delta,J}(P,D_Z)G_{\Delta,J}(X,P;W,Z),
\end{equation}
in which the higher spin fields and the bulk-to-boundary propagator are both written in terms of polynomials. As for the $C_{\Delta,J}(P,D_Z)$, we can treat it as an operator polynomial $C^{A_1\cdots A_J}_{\Delta,J}D^Z_{A_1\cdots A_J}$ thus determining the matching condition for $H^{A_1\cdots A_J}$ is equivalent to determining the condition for $C_{\Delta,J}^{A_1\cdots A_J}$.\\ \\
 \textbf{i) Spin $J=1$ Match}\\ \\
 For the spin 1 case, we first write down the expansion of fields in terms of the polynomial
 \begin{equation}
     H(X,W)=\frac{1}{\frac{d}{2}-1}\;\int dP \;C^A(P)\;D^Z_A \;G_{\Delta,J=1}(X,P;W,Z),
 \end{equation}
 in which $C^A$ are coefficients carrying the tensor indexes. After substituting $G_{\Delta,J}$, $D^Z_A$ into the integrand and applying $K_A^W$ on the polynomial, we obtain the tensor in embedding space
 \begin{equation}
     H_A(X)=\int dP\;\frac{-2(P\cdot X)C_A+2(C\cdot X)P_A}{(-2P\cdot X)^{\Delta+1}}
     \label{ts}
 \end{equation}
 in which we can check that the transverse condition $X\cdot H=0$ is automatically satisfied for arbitrary $C^A(P)$. Moreover, one can also directly see from the above expression that there is a redundancy of the coefficients
 \begin{equation}
     C^A(P)\longrightarrow C^A(P)+\lambda P^A,
 \end{equation}
 which tells us a shift of the coefficient by $\lambda P^A$ will give us the same higher spin field. We call this the pure gauge of coefficients since it works the same way as the pure gauge for CFT spin fields and later we will see that the consistence of matching conditions requires us to fix the pure gauge.\\ \\ Now, if we just consider the integrand, the matching condition \eqref{hsm1} tells us that
 \begin{eqnarray}
     -(P\cdot X)_E\; C^E_a+(C^E\cdot X)_E\;P_a=-(P\cdot X)_L\;C_a^L+(C^L\cdot X)_L\;P_a,
 \end{eqnarray}
in which $C^E_A$, $C^L_A$ are the tensor coefficients in Euclidean and Lorentzian signature and they have already been pulled back to the AdS space thus labelled by $C^E_a$, $C^L_a$. Here we should note that the above equation is restricted on the surface $\Sigma$ which we did not write down explicitly for short. Since $P_a$ serve as variables in the integrand, to solve the above matching condition for arbitrary $P$, we should make each term on both sides fit. Furthermore, noting that $(P\cdot X)_E=(P\cdot X)_L$ is guaranteed by the rotation of embedding coordinates, therefore the non trivial conditions are determined as
\begin{eqnarray}
C_a^E|_\Sigma=C_a^L|_\Sigma, \qquad (C^E\cdot X)_E|_\Sigma=(C^L\cdot X)_L|_\Sigma \label{solution}. 
\end{eqnarray} 
The first equation is just the statement that the coefficients are continuous at the joint surface $\Sigma$ in $\rm AdS$ spacetime and the second one can be simplified to $iC^E_0=C^L_0$, which is the feature of Wick rotation on   coefficient tensor fields associated with the rotation of embedding space coordinates.\footnote{Here $C_{A=0}$ represents the tensor in embedding space while later we will meet the zero component in $AdS$ space and we denote it as $C_{a=0}$.}\\ \\
Here, we should stop to check that if the two conditions in \eqref{solution} could be compatible. Actually, if we consider the coordinate transformation of time direction under the condition $\tau=iT, t=T$ and $C^E_0=iC_0^L$, the relation we get in $\rm AdS$ spacetime should be $C_{a=0}^E=iC_{a=0}^L$  at the joint surface rather than the first one in \eqref{solution} $(C^E_{a=0}=C^L_{a=0})$, which implies in fact the matching condition we should impose on the tensor field is that
\begin{eqnarray}
    &&H_{a}^{E}(X)|_\Sigma=H_{a}^{L}(X)|_\Sigma, \qquad{\rm for} \qquad a\neq 0\\\nonumber \\
    &&H_{0}^{E}(X)|_\Sigma=iH_{0}^{L}(X)|_\Sigma \label{spm}.
\end{eqnarray}
And the modification of the matching condition in time direction results from the convention when we are doing the calculation while we can treat it as the feature of the rotation of vector fields, which we do not need to consider in scalar field matching.\\ \\
Physically, suppose that we have the higher spin action in the hand and try to match actions of different signatures together, the special matching condition  \eqref{spm} for time component will make the invariant term like 
\begin{equation}
    \eta^{ab}H_a^LH_b^L=\delta^{ab}H_a^EH_b^E
\end{equation}
joint smoothly at the surface $\Sigma$. Thus now we can see that the condition \eqref{spm} becomes obvious from physics point of view even though it was derived by checking the consistence of the propagator matching. Next, we come to study the second matching equation. Based on the above discussion, we determine the modified matching conditions as
\begin{eqnarray}
&&\partial_\tau H^E_a(X)|_\Sigma+i\partial_t H^L_{a}(X)|_\Sigma =0, \qquad {\rm for} \qquad a\neq 0\\ \nonumber\\
&&\partial_\tau H^E_0(X)|_\Sigma-\partial_t H^L_{0}(X)|_\Sigma =0,
\end{eqnarray}
in which we have taken the rotation of $H_0$ into consideration.  As for the derivative with respect to time, we need to substitute \eqref{de} and \eqref{dl} into \eqref{ts} and then obtain
  \begin{eqnarray}
     \partial_\tau  H_A(X)&=&\int dP\;\frac{-2(-2X^0P_-+X^+P_0)C_A+2(-2X^0C_-+X^+C_0)\;P_A}{(-2P\cdot X)_E^{\Delta+1}}\nonumber\\\nonumber  \\&+&\frac{2(\Delta+1)(-2X^0P_-+X^+P_0)(-2(X\cdot P)_EC_A+2(C\cdot X)_EP_A)}{(-2X\cdot P)_E^{\Delta+2}} \end{eqnarray}
for Euclidean signature and 
   \begin{eqnarray}
     \partial_t  H_A(X)&=&\int dP\;\frac{-2(2X^0P_-+X^+P_0)\;C_A+2(-2X^0C_-+X^+C_0)\;P_A}{(-2P\cdot X)_L^{\Delta+1}}\nonumber\\\nonumber \\&+&\frac{2(\Delta+1)(2X^0P_-+X^+P_0)(-2(X\cdot P)_L\;C_A+2(C\cdot X)_L\;P_A)}{(-2X\cdot P)_L^{\Delta+2}}
 \end{eqnarray}
 for Lorentzian signature. It will be convenient to note the differences between them mainly come from the inner product $(\cdot)_{L/E}$ and the sign in front of $\pm X^0P_-$. Then, using the matching condition \eqref{hsm2}, we have
 \begin{eqnarray}
 &&\Delta(-X\cdot P)_E\;(-2X^0P_-+X^+P_0)C_a +((-X\cdot P)_E\;(-2X^0C_-+X^+C_0)\nonumber\\\nonumber\\&&+(\Delta+1)(C\cdot X)_E\;(-2X^0P_-+X^+P_0))P_a +i(\cdots)_L=0,
 \end{eqnarray}
 in which $a\neq 0$ and $(\cdots)_L$ represents the Lorentzian version of the formula we wrote down explicitly and one can check that the above equation is trivial provided the conditions in \eqref{solution} are satisfied. The new restriction comes from the study of the zero component for $a=0$, taking the extra contribution $2X^+H_-$ into consideration, we have
 \begin{eqnarray}
 &&\Delta(-X\cdot P)_E\;(-2X^0P_-+X^+P_0)C_{a=0}+(P\cdot X)_E^2\;C_-X^+-(P\cdot X)_E(C\cdot X)_EP_-X^+ \nonumber\\\nonumber\\&&+((-X\cdot P)_E\;(-2X^0C_-+X^+C_0)+(\Delta+1)(C\cdot X)_E\;(-2X^0P_-+X^+P_0))P_0\nonumber \\\nonumber\\&&-(\cdots)_L=0.
 \end{eqnarray}
 To solve it, one should impose the condition
 \begin{equation}
     C_-^E|_\Sigma=C_-^L|_\Sigma, \label{gauge}
 \end{equation}
 which gives us the proper gauge of the embedding coordinates on the joint surface $\Sigma$. To understand such an extra condition, first we consider the degrees of freedom of the matching conditions in terms of the coefficients $C^A$. Since we are dealing with $d+1$ dimensional AdS spacetime, there are $d+1$ equations for us to solve in \eqref{solution}, together with the gauge condition \eqref{gauge}, we have $d+2$ matching conditions, which uniquely fix the $d+2$ coefficients $C^A$ in embedding space. We can also understand this in a way that is similar to the direct match of $H_A$. As we have mentioned before, after imposing the transverse condition, there will be a one to one correspondence between the fields in the embedding space and the fields on the AdS surface. But the transverse conditions will not introduce any restriction on the coefficients therefore we have the pure gauge redundancy. Now we see that the pure gauge should be fixed when doing the matching.
 \\\\
 \textbf{ii) Spin $J=2$ Match}\\\\
 For the spin 2 field, we just consider the symmetric and traceless tensor field $H_{AB}$ for simplicity. In this case, the polynomial can be written as
 \begin{equation}
     H(X,W)=\frac{2}{(d+1)(d-1)}\;\int dP \;C^{AB}(P)\;D^Z_A D_B^Z \;G_{\Delta,J=2}(X,P;W,Z),
 \end{equation} 
 which corresponds to the tensor
 \begin{eqnarray}
     H_{AB}(X,W)=\int dP\;\frac{4}{(-2X\cdot P)^{\Delta+2}}\;((X\cdot P)^2 C_{AB}\label{ten2} \\\nonumber\\+P_AP_BX^CX^DC_{CD}-(X\cdot P)P_AX^CC_{CB}-(X\cdot P)P_BX^CC_{CA})\nonumber.
 \end{eqnarray}
After imposing the matching condition, we have
 \begin{eqnarray}
     (X\cdot P)_E^2 \;C^E_{ab}+P_aP_bX^cX^dC^E_{cd}-(X\cdot P)_E\;P_a\;X^cC^E_{cb}-(X\cdot P)_E\;P_b\;X^cC^E_{ca}\nonumber\\\nonumber\\= (X\cdot P)_L^2 \;C^L_{ab}+P_aP_bX^cX^dC^L_{cd}-(X\cdot P)_L\;P_a\;X^cC^L_{cb}-(X\cdot P)_L\;P_b\;X^cC_{ca}^L,
 \end{eqnarray}
 in which the tensor $C_{AB}$, $P_A$, $X_A$ together with the inner product have been pulled back to the $AdS$ Poincar$\acute{{\rm e}}$ coordinates. Similar to the study of spin $J=1$ case, we deduce the solution to be 
\begin{eqnarray}
&&C_{ab}^E|_\Sigma=C_{ab}^L|_\Sigma \quad {\rm for}\quad a\neq0, \qquad C^E_{0b}|_\Sigma=iC^L_{0b}|_\Sigma\quad {\rm for } \quad a=0,\\\nonumber\\&& C^E_{ab} \;X^a|_\Sigma=C^L_{ab}\; X^a|_\Sigma,
\label{res}
\end{eqnarray}
in which the first implies the continuation of the tensor field, the second corresponds to the Wick rotation on the time direction and the third implies the match of inner product, and we note that they are compatible by imposing $C^E_{0B}=iC^L_{0B}$.\\\\
Now, we come to study the second matching condition and check if the restrictions showed in \eqref{res} are enough. Substituting \eqref{de} into \eqref{ten2}, we see that the derivative with respect to the Euclidean time is given by
 \begin{eqnarray}
&&\partial_\tau H_{AB}(X,W)=4\int dP\;\frac{2}{(-X\cdot P)_E^{\Delta+2}}((-2X^0P_-+X^+P_0)((X\cdot P)_E\;C_{AB} \\\nonumber\\
     &&-P_{(A}X^CC_{CB)})+P_AP_BX^D(-2X^0C_{-D}+X^+C_{0D})-(X\cdot P)_EP_{(A}(-2X^0C_{-B)}+X^+C_{0B)}))\nonumber\\\nonumber\\&&+\frac{2(\Delta+2)(-2X^0P_-+X^+P_0)}{(-2X\cdot P)_E^{\Delta+3}}\;((X\cdot P)_E\; C_{AB}+P_AP_BX^CX^DC_{CD}-2(X\cdot P)_E\;P_{(A}X^CC_{CB)}).\nonumber
 \end{eqnarray}
 in which we use the convention for symmetrising the free tensor index $C_{(AB)}=\frac{1}{2}(C_{AB}+C_{BA})$. And, as for the Lorentzian time derivative, all we need to do is to change the sign in front of $X^0P_-$ and the notion of inner product while we will not show that explicitly here. Consider the integrand, we will obtain
 \begin{eqnarray}
&&P_aP_b((\Delta+2)(-2X^0P_-+X^+P_0)X^cX^dC_{cd}+(-2P\cdot X)_E X^d(-2X^0C_{-d}+X^+C_{0d}))\nonumber\\\nonumber\\&&-P_{(a}(X\cdot P)_E(\Delta(-2X^0P_-+X^+P_0)X^cC_{cb)}+(-2P\cdot X)_E\;(-2X^0C_{-b)}+X^+C_{0b)}))\nonumber\\\nonumber\\
     &&+\Delta(-2X^0P_-+X^+P_0)(X\cdot P)_E^2\;C_{ab}+i(\cdots)_L=0
 \end{eqnarray}
for $a,b\neq 0$ from the second the matching condition, which will be trivial provided the restrictions we imposed in \eqref{res} are satisfied. After all, to make the matching condition fits at the $a=0$ and $b\neq0$ direction, one should also impose the gauge constraint
\begin{equation}
    C^E_{-b}|_\Sigma=C^L_{-b}|_{\Sigma},
\end{equation}
which results from the matching condition
\begin{equation}
    \partial_\tau H^E_{0b} - \partial_t H^L_{0b}=0.
\end{equation}
\subsection{Embedding the BTZ }\label{btzs}
\noindent
Based on the study of match conditions on the AdS background, we now come to a more interesting case which is the BTZ black hole. The treatment of BTZ black hole is similar to the $\rm AdS_3$ since they are locally isomorphic  therefore all the results for AdS could be generalised to the BTZ black hole.  Also, as we have mentioned, there is not a unique way to lift a AdS solution to the entire embedding space. In this section, we will use BTZ solutions as an example to illustrate more detail on this therefore achieve a bettering understanding on how to deal with AdS solutions in embedding space or vice versa.\\ \\
First, we will introduce BTZ black hole \cite{Banados:1992gq,Banados:1992wn} in embedding space via identifying points along the trajectory generated by the chosen Killing vector in ${\rm AdS_3}$. We start from the embedding of Lorentzian $\rm AdS_3$ given by
\begin{equation}
    -(X^{-1})^2-(X^0)^2+(X^1)^2+(X^2)^2=-R^2,
\end{equation}
in which the $SO(2,2)$ symmetry is manifested and $R$ is related to the cosmological constant by $-\Lambda=R^{-2}$. The $\rm AdS_3$ in embedding space has the metric 
\begin{equation}
    ds^2=-(dX^{-1})^2-(dX^0)^2+(dX^1)^2+(dX^2)^2,
\end{equation}
and in order to get the black hole geometry, we introduce the Killing vector
\begin{equation}
    \xi=\frac{r_+}{R}\left(X^{-1}\frac{\partial}{\partial X^2}+X^{2}\frac{\partial}{\partial X^{-1}}\right),\label{killing}
\end{equation}
where $r_+$ is a constant characterising the size of horizon. Given the initial point $P$, the Killing vector will generate a curve $c(t)=e^{t\xi}P$ in which $\xi$ serves as the tangent vector along $c(t)$. Here, we are going to identify the points on the curve such that $t\in 2\pi \mathbb{Z}$ and those points which are invariant under the transformation $e^{2\pi \mathbb{Z}\xi}$ will become singularities of the quotient space ${\rm AdS_3}\setminus\sim$. Although the quotient space satisfies the Einstein equation at the regular points, we still need to get rid of the closed timelike curves, resulting from the identification procedure, to obtain a reasonable causal structure. According to the property of the Killing field, $\xi\cdot \xi$ is preserved along the curve $c(t)$, thus the necessary condition for the absence of closed timelike curves is $\xi^2>0$ everywhere in the manifold and in terms of the embedding coordinates, we have
\begin{equation}
    (X^{-1})^2-(X^2)^2=X^+X^->0,
\end{equation}
which gives us the black hole geometry with a proper causal structure. To see the effect of identification, we will introduce the $(t,r,\phi)$ coordinate defined as
\begin{eqnarray}
X^{-1}&=&\frac{Rr}{r_+}\;{\rm cosh}(\frac{r_+}{R}\phi),\label{cor1}\\\nonumber \\
X^{2}&=&\frac{Rr}{r_+}\;{\rm sinh}(\frac{r_+}{R}\phi),\label{cor2}\\\nonumber \\
X^0&=&\frac{R}{r_+}\sqrt{r^2-r_+^2}\;{\rm sinh}(\frac{r_+t}{R^2}),\\\nonumber \\
X^1&=&\frac{R}{r_+}\sqrt{r^2-r_+^2}\;{\rm cosh}(\frac{r_+t}{R^2}),
\end{eqnarray}
in which $-\infty<t,\phi<\infty$, $r\ge r_+$, and for $0\le r\le r_+$ the expressions for $X^{-1}$, $X^2$ are the same while we have
\begin{eqnarray}
    X^0&=&\frac{-R}{r_+}\sqrt{r_+^2-r^2}\;{\rm cosh}(\frac{r_+t}{R^2}),\label{in1}\\\nonumber \\
X^1&=&\frac{-R}{r_+}\sqrt{r_+^2-r^2}\;{\rm sinh}(\frac{r_+t}{R^2}),\label{in2}
\end{eqnarray}
for $X^0$, $X^1$. These two patches together will cover the  $\rm AdS_3$ and Killing vector \eqref{killing} becomes
\begin{equation}
    \xi =\frac{\partial}{\partial\phi}
\end{equation} 
once we have pushed it back to the hyperboloid described by $(t,r,\phi)$. Therefore, in the new coordinates, the identification of points under $e^{2\pi\mathbb{Z} \xi}$ is equivalent to imposing $\phi\cong \phi+2\pi$ and then we obtain the black hole metric
\begin{equation}
    ds^2=-\left (\frac{r^2-r_+^2}{R^2}\right )dt^2+\left(\frac{R^2}{r^2-r_+^2}\right)dr^2+\frac{r^2}{R^2}d\phi^2,
    \label{BTZ m}
\end{equation}
from which we can see that the horizon lies at $r=r_+$, thus in embedding coordinates, the horizon is
\begin{equation}
    X^+X^-=R^2.
\end{equation}
For simplicity, we usually choose to perform the coordinates transformation
\begin{equation}
    t\longrightarrow \frac{r_+}{R^2}t,\qquad r\longrightarrow \frac{r}{r_+},\qquad \phi \longrightarrow \frac{r_+}{R}\phi
\end{equation}
and get
\begin{equation}
    ds^2=-(r^2-1)dt^2 +\frac{dr^2}{r^2-1}+r^2d\phi^2,
    \label{BTZ metric}
\end{equation}
with the periodic condition $\phi=\phi+\frac{2\pi r_+}{R}$.\\ \\ Now, we come to introduce the solution of the Kelin-Gordon equation for the scalar field $\Phi$ on the BTZ black hole background \eqref{BTZ m}, written as
\begin{equation}
     \nabla^2\Phi_{\Delta}-m^2\Phi_{\Delta}=0,
     \label{kg}
\end{equation}
in which $\nabla$ represents the Laplacian operator on the curved spacetime $G$. The detail of the solution is shown in the Appendix \ref{btz} and here we just discuss the results. As for the solution near the horizon $r=r_+$, there are two independent modes
\begin{equation}
    \psi_\pm=e^{ \frac{ir_+}{R^2}(\pm \omega t-kR\phi)}f_{\Delta}(\pm\omega,k,\frac{r}{r_+}),
\end{equation}
from which we can see that the behaviour of the modes near the horizon depends on the frequency $\omega$ which characterizes the propagation of the modes along the circle, called left or right moving modes. If we consider the solution at the spatial infinity making $r\to \infty$, the modes now become
\begin{equation}
     \psi_\pm=e^{\frac{ir_+}{R^2}( \omega t-kR\phi})f_{\Delta^\pm}(\omega,k,\frac{r}{r_+}),
\end{equation}
where the scale dimension is given by the dictionary $\Delta(\Delta-d)=m^2$ treated as a constant parameter independent of the spacetime coordinates. We can see that these two modes will behave like $r^{-\Delta}$ and $r^{\Delta-2}$ asymptotically, corresponding to the source and normalizable modes, which have been studied in the vacuum $ \rm AdS_3$ case.\\ \\
It is worthwhile to note that, although we obtain four different kinds of modes in total, this dose not mean that the scalar field $\Phi$ should be the linear combination of these four modes. The reason that we obtain four modes here is that we are expanding the same function around different singular points thus the basis changes. Near horizon, the basis carries the information of the direction of the propagation while, at the infinity, the basis carries the information of asymptotic behavior of the field according to the radius $r$. 
\subsubsection*{Embedding the Solution}
In the above section, we have studied the embedding structure of the BTZ black hole and obtained the solution of Klein-Gordon equation on AdS background while in this section we are going to lift the solution from the AdS hyperboloid to the embedding space. \\ \\
There are various ways of extending a function on a hyperboloid to the embedding space. Since we can foliate the embedding space with AdS surfaces of different radii $R$, a natural embedding way is just treating the radii $R$ as a variable and extending the solution obtained from some standard surface $R=R_0$ to the surfaces with different radii $R$. This is easy to do mathematically while we should note that the extended function will not be the solution of Kelin-Gordon equation on other AdS surfaces except for the surface $R_0$. Vice versa, to fully retain the physical meaning of the extended function, one can solve the KG equation \eqref{kg} on each surface but it would be hard to smoothly glue them together.\\ \\
Here, instead of  considering the extension of the solution directly, we consider the extension of the KG equation. Using the generator of the $SO(2,2)$ group
\begin{equation}
    J_{AB}=X_A\frac{\partial}{\partial X^B}-X_B\frac{\partial}{\partial X^A},
\end{equation}
we can construct the quadratic Casimir directly in embedding space \cite{2016}
\begin{equation}
    \frac{1}{2}J_{AB}J^{BA}\Phi_{\Delta}(X)=R^2\nabla^2_{AdS}\;\Phi_{\Delta}(X),
\end{equation}
in which $\nabla_{AdS}=\nabla$ is the Laplacian operator for the BTZ black geometry which is locally isometric to $\rm AdS_3$. The above expression can be treated as the decomposition of an embedding operator along the AdS surfaces. Given the quadratic Casimir, the equation of motion in embedding space is deduced to be
\begin{equation}
    \frac{1}{2}J_{AB}J^{BA}\Phi_\Delta(X)=m^2R^2\Phi_\Delta(X),
    \label{emq}
\end{equation}
from which we see that the mass term $m^2R^2$ now depends on the radii of the surfaces, and it will be reduced to the KG equation at radii $R=1$. Since $[J_{AB},\;X^2+R^2]=0$, we can solve the Equation \eqref{emq} on each surface, following the same method of solving KG equation, provided we make
\begin{equation}
    m^2R^2=\Delta_{R}(\Delta_R-2).
\end{equation}
Now, if we write solution in terms of the $(r,t,\phi,R)$ coordinates i.e, $\Phi(X)=\Phi(t,r,\phi,R)$, the modes will become
\begin{equation}
    \psi_\pm=e^{ \frac{ir_+}{R^2}(\pm \omega t-kR\phi)}f_{\Delta_R}(\pm\omega,k,\frac{r}{r_+})
\end{equation}
at the horizon while, at the infinity, the two modes are
\begin{equation}
     \psi_\pm=e^{\frac{ir_+}{R^2}( \omega t-kR\phi)}f_{\Delta_R^\pm}(\omega,k,\frac{r}{r_+}),
\end{equation}
in which we not only make the phase part $R$ dependent but also the scale dimension $R$ dependent. These solutions are smooth in embedding space and they are also solutions of the scalar KG equation on each AdS surface with mass $mR$, i.e, now the equation becomes
\begin{equation}
    \nabla^2\Phi_{\Delta_R}-m^2R^2\Phi_{\Delta_R}=0,
\end{equation}
which can be understood that we are considering excitation spectra or fluctuation of particle with mass of order $R$ on each AdS surfaces.
\subsection*{BTZ Propagator}
In this part, based on the study of the geometry of BTZ black holes, we will generalise the vacuum AdS propagators to the black hole case and discuss the thermal feature of the boundary theories. Using these propagators, following the same procedure, one can write down on shell higher spin fields propagating on the black hole background therefore obtain the match conditions. It turns out that the match condition will have the same form as the ${\rm AdS_3}$ case since the local property of the propagator remain the same.\\ \\
As we have already known, BTZ geometry is obtained from $\rm AdS_3$ by the identification of the points $\phi\cong \phi+2\pi$. In the embedding coordinates, using \eqref{cor1} and \eqref{cor2}, one can deduce that this is equivalent to 
\begin{equation}
    X^{\pm}\cong e^{\pm 2\pi r_+} X^{\pm}.
\end{equation}
Here, to study the identified points more carefully, we introduce the notion
\begin{equation}
      \widetilde{X}^n:=(e^{+2n\pi r_+}X^+,\;e^{-2n\pi r_+}X^-\;X^0,\;X^1),
\end{equation}
in which we use the superscript $n$ to represent the winding number of the coordinates. Moreover, we should note that the points $\widetilde{X}^n$ are distinguishable in the $\rm AdS_3$ geometry while they form a cover of a single point  $X$ of BTZ black hole
\begin{equation}
    X=\widetilde{X}^0\cong \widetilde{X}^1\cdots   \widetilde{X}^n\cong  \widetilde{X}^{n+1}\cdots.\label{eq}
\end{equation}
Following such convention,  one can directly write down the BTZ boundary-bulk propagator
\begin{equation}
  G^{BTZ}_{\Delta,J}(X,P;W,Z)=\sum_{n=-\infty}^{\infty}\frac{((-2P\cdot \widetilde{X}^n)(W\cdot Z)+2(W\cdot P)(Z\cdot \widetilde{X}^n))^J}{(-2P\cdot \widetilde{X}^n)^{\Delta+J}}
\end{equation}
with the help of the method of images introduced in \cite{1999,Kraus:2002iv}. Basically, the infinite sum over the winding number on the right hand side is used to construct a function that is invariant at the points $\widetilde{X}^n$ so that the relation \eqref{eq} is manifested in the context of the BTZ geometry.\\ \\
To see how this works in a more specific way, we go back to the $(t,r,\phi)$ coordinates. The bulk points $X$ are well defined in the previous section and here for the boundary points $P$, we write them as
\begin{equation}
    P=(P^{+},\;P^-,\;P^0,\;P^1)=(e^{r_+\phi'},\;e^{-r_+\phi'},\;{\rm sinh}(r_+t'),\;{\rm cosh}(r_+t')),
\end{equation}
in which the light ray condition $P^2=0$ is satisfied and the $(t',\phi')$ are in fact coordinates of a cylinder. With these coordinates, for the scalar case, one can write the bulk boundary operator as
\begin{equation}
    G^{+}_{\Delta,0}(X,P)=\sum _{n=-\infty}^{\infty}\frac{1}{\left(-\frac{\sqrt{r^2-r_+^2}}{r_+}{\rm cosh}(r_+\delta t)+\frac{r}{r_+}{\rm cosh }(r_+(\delta\phi+2\pi n)) \right)^\Delta}
\end{equation}
for $r>r_+$ and $\delta\phi=\phi-\phi'$, $\delta t=t-t'$. This is the bulk-boundary propagator when the bulk point is outside the horizon. For the inside horizon propagator, we should use the coordinate \eqref{in1}, \eqref{in2} and then obtain
\begin{equation}
    G^{-}_{\Delta,0}(X,P)=\sum _{n=-\infty}^{\infty}\frac{1}{\left(-\frac{\sqrt{r_+^2-r^2}}{r_+}{\rm sinh}(r_+\delta t)+\frac{r}{r_+}{\rm cosh }(r_+(\delta\phi+2\pi n)) \right)^\Delta}.
\end{equation}
For the boundary correlation functions, the polynomial of higher spin two point function 
\begin{equation}
   \langle \mathcal{O}_J(P_1)\mathcal{O}_J(P_2)\rangle(Z_1,Z_2)=\sum_{n=-\infty}^{\infty}\frac{((-2 \widetilde{P}^n_1\cdot P_2)(Z_1\cdot Z_2)+2(\widetilde{P}_1^n \cdot Z_2)(P_2\cdot Z_1))^J}{(-2\widetilde{P}_1^n\cdot P_2)^{\Delta+J}}
\end{equation}
is obtained by the projection of the bulk-boundary propagator to the $r^\Delta$ term, in which $\widetilde{P}^n$ are defined as
\begin{equation}
    \widetilde{P}^n:=(e^{+2n\pi r_+}P^+,\;e^{-2n\pi r_+}P^-,\;P^0,\;P^1)
\end{equation}
thus the temperature $T=r_+/2\pi$ can be deduced after going to the Euclidean signature $t=-i\tau$.
\newpage
\section{General Principles of Lorentzian Continuation} \label{continuation}
In this section we will do the Wick rotation of CFT completeness relation by matching the scale dimension and spin in different signatures properly then present a Lorentzian AdS/CFT map based the derivation of Euclidean AdS/CFT map in \cite{Aharony:2020omh,Aharony:2021ovo}. Given the Lorentzian AdS/CFT map, we will present the matching conditions for the quadratic action then obtain the matching conditions for higher order action.
\subsection{Scale Dimension and Spin}
\begin{figure}
\begin{tikzpicture}
\tikzset{decoration={snake,amplitude=.4mm,segment length=2mm,post length=0mm,pre length=0mm}}
\draw[thick,->](-2.5,-2)--(-2.5,2);
\draw[thick,->](-3.5,0)--(3.5,0);
 \draw[very thick,red](0,-2)--(0,2);
   \draw (0,-2.5) node{\small{$\Delta=\frac{d}{2}+is$}};
   \draw(-1,-0.5)node{$\Delta_0$};
 \draw[very thick,blue](-1,0)--(3.5,0); 
  \filldraw [black] (-1,0) circle (2pt);
    \draw (3,1.75) node{$\Delta$};
    \draw [thick,black,->](0.2,1.5) .. controls (1.4,1.4) .. (2,0.2);
    \draw[thick](2.75,1.5)--(3.25,1.5);
    \draw[thick](2.75,1.5)--(2.75,1.9);
\end{tikzpicture}
\hspace{3.0em}
\begin{tikzpicture}
\tikzset{decoration={snake,amplitude=.4mm,segment length=2mm,post length=0mm,pre length=0mm}}
\draw[thick,->](-2.5,-2)--(-2.5,2);
\draw[thick,->](-3.5,0)--(3.5,0);
 \draw[very thick,red](0,-2)--(0,2);
   \draw (0,-2.5) node{\small{$\Delta=\frac{d}{2}+is$}};
   \draw(-1,-0.5)node{$\Delta_0$};
   \draw[thick,->](-1,0.2)--(-1.5,0.2);
    \draw[very thick,blue](-1,0)--(3.5,0); 
  \filldraw [black] (-1,0) circle (2pt);
    \draw (3,1.75) node{$\Delta$};
    \draw[thick](2.75,1.5)--(3.25,1.5);
    \draw[thick](2.75,1.5)--(2.75,1.9);
    \draw[very thick,red, <-] (2.3,0.5) arc (0:360:0.3);
      \filldraw [black] (2,0.5) circle (1pt);
      \draw[thick,->](2,0.5)--(1.5,0.5);
\end{tikzpicture}
\caption{The analytic continuation of the scale dimension is shown on the left hand in which principal series lies on the red line and physical scale dimension distributes along the blue line on the real axes, with the lower bound $\Delta_0$. In the right figure, the motion of the poles according to the changing of the lower bound $\Delta_0$ is illustrated. In order to eliminate such effects caused by the motion of poles, an extra part of the contour winding around the singularity should be added into the original contour $\gamma$. The new one is denoted as $\gamma_J$ and illustrated by the red curves. }
\label{f3}
\end{figure}
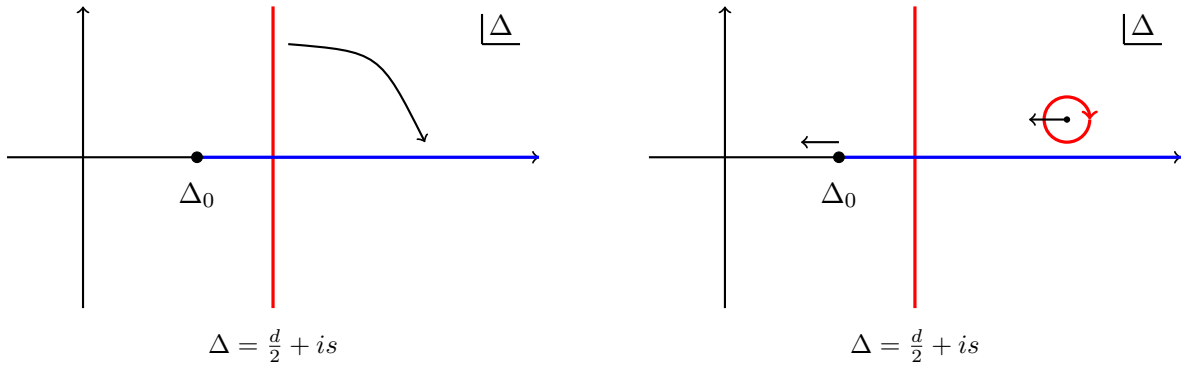

First we will discuss the behavior of scale dimension $\Delta$ and spin $J$ in different signatures from both the conformal field theory and gravity theory point of view, which plays the central role in the context of Wick rotation and AdS/CFT matching.
\begin{table}[H]
\centering
\begin{tabular}{|c|c|p{1.6cm}|c|}
\hline
 & Symmetry Group & Principal Series & Parameter\\
\hline &&&\\
Euclidean & $SO(d+1,1)\sim SO(1,1)+SO(d)$ & $\quad\mathcal{E}_{\Delta,\;J}$&$\Delta=\frac{d}{2}+i\mathbb{R}$,\quad$J\in \mathbb{Z}$\\
&&&\\
\hline &&&\\
 & $SO(d,2)\sim SO(1,1)+$ & &$\Delta=\frac{d}{2}+i\mathbb{R}$\\ Lorentzian&&$\;\mathcal{P}_{\Delta,\;J,\;\lambda}$& \\ &$SO(1,1)+SO(d-2)$&&$J=\frac{d-2}{2}+i\mathbb{R}$,\;\;$\lambda\in \mathbb{Z}$\\&&& \\
\hline
\end{tabular}
\caption{Harmonic analysis of the conformal symmetry for $d$ spacetime dimensional Euclidean and Lorentzian conformal field theory. The principal series for Euclidean and Lorentzian signature are labelled by $\mathcal{E}$ and $\mathcal{P}$, respectively.}
\label{t1}
\end{table}
\noindent We start from the study of representation theory of the conformal symmetry group. As for the scale dimension induced by the dilaton operator $D$, it generates a noncompact direction, i.e noncompact subgroup $SO(1,1)$, in both signatures and takes the value $\Delta=\frac{d}{2}+i\mathbb{R}$ on the same principal series provided the representation is unitary while the story for the spin is different.\\ \\
For the spin in Euclidean signature, it represents the compact group $SO(d)$ thus takes the integer value $J\in\mathbb{Z}$ and in Lorentzian signature it represents the noncompact time direction generated by the operator $M_{01}$ therefore becomes continuous on the principal series $J=\frac{d-2}{2}+i\mathbb{R}$. The decomposition of the symmetry group and the range of parameters are summarized in Table.\ref{t1}. We should note that in Lorentzian signature we usually take the representation of compact group $SO(d-2)$ to be trivial therefore make $\lambda=1$. In the end, to get the physical operators, we need to analytically continue the $\Delta, J$ from principal series to real axes while the rotation of the scale dimension is shown in LHS of Fig.\ref{f3}. \\\\\
Practically, in order to get an analytic function, the contour along the principal series is not enough and we need to deform the contour especially when we meet poles in the complex plane. For example, in the completeness relation \eqref{come} or \eqref{leq}, $\Delta_0=\frac{d}{2}$ is defined on the principal series. But physically, as the lower bound of the scale dimension, $\Delta_0$ itself depends on the theory thus it varies along the real axes according to the interaction picture we have. Note that $\Delta_0$ serves as a parameter in the constant factor thus the position of the pole of coefficients $N_{\Delta,J}$ \footnote{Given the complete basis which goes through the principal series of $\Delta,J$, the coefficients $N_{\Delta,J}$ for arbitrary field configuration are automatically determined after performing the mode expansion. More careful study will be shown later.} will move when we change the value of $\Delta_0$. Taking these effects into consideration, if one wants to extend the equation apart from $\Delta=\frac{d}{2}$ analytically, extra contours going around the poles should be added properly, illustrated in RHS of Fig.\ref{f3}, and the Euclidean case is discussed in \cite{Aharony:2020omh}. To distinguish the deformed contours from the original one $\gamma$, we denote them as $\gamma_E$ and $\gamma_L$  corresponding to the Euclidean and Lorentzian CFT \footnote{In fact we should impose that $\gamma_E=\gamma_L$ to make fields satisfy the match condition and we label them together as $\gamma_J$, which means that the deformation of the principal series depends on the spin $J$. }.\\ \\
The above discussion comes from the study of the symmetry group on field theory side, which must have implications in the bulk theory with gravity due to the correspondence. The connections come in when we try to construct a map between the bulk fields and the boundary fields via the decomposition \eqref{de1} and \eqref{de2}. In \eqref{de1} and \eqref{de2}, we just state without explanation that $\Delta$ should lie on the principal series even though harmonic analysis does not apply to the gravity theory. Here we point out that if one wants to complete the map between the bulk and boundary field, one should choose the contour $\gamma_E$ in \eqref{de1}
 and the contour $\gamma_L$ in \eqref{de2} and such choice of contour in the bulk theory can be interpreted as selecting proper modes of the bulk-boundary propagator, i.e, we have the correspondence
 \begin{equation}
     {\rm Poles\;\; of}\;\;N_J(\Delta) \;,\;N(\Delta,J)\qquad\longleftrightarrow \qquad {\rm Bulk\;\; Modes },
 \end{equation}
 where $N_J(\Delta)$ and $N(\Delta,J)$ are the coefficients of the bi-local fields given the conformal basis labelled by the parameter $J$ and $\Delta$. This is similar to the case we have met in the study of quasinormal modes \cite{birmingham2001choptuik,birmingham2002conformal,birmingham2003relaxation} of BTZ black holes in which the pole of frequency in the BTZ solutions is related to the pole of boundary CFT correlation functions. In the study of SYK model \cite{Maldacena:2016upp,maldacena2016remarks,Polchinski:2016xgd,Kitaev:2017awl}, four point functions are also expanded by the conformal basis consisting of continuous modes on the principal series together with discrete modes resulting from poles of $N_J(\Delta)$ and such discrete mode in the bulk corresponds Goldstsone boson mode called dilaton that describes the symmetry broken of $\rm AdS_2$. In the context of HS/CFT duality, we will use three point correlation functions as the basis of the boundary bilocal fields and it turns out the both of the physical and ghost modes in the bulk are generated by the boundary discrete modes.\\ \\
 Having discussed the role of scale dimension, we now come to the study of spin $J$ while it will become more complicated even though we just stay on the CFT side. The subtlety firstly comes in when we try to analytically extend the spin $J$ in Lorentzian signature apart from the principal series since the physical spin are discrete integer numbers \cite{mack1977all} on real axes rather than a continuous interval and there is no mechanism telling us how the basis collapse or whether the physical basis is still complete.\\ \\
  \begin{figure}
\begin{tikzpicture}
\tikzset{decoration={snake,amplitude=.4mm,segment length=2mm,post length=0mm,pre length=0mm}}
\draw[thick,->](-2.5,-2)--(-2.5,2);
\draw[thick,->](-3.5,0)--(3.5,0);
 \filldraw [black] (-1.5,0) circle (2pt);
\filldraw [black] (-2.5,0) circle (2pt);
\draw[ very thick,blue](-2.5,-0.3)--(-1.5,-0.3);
\draw[ very thick,blue,->](-0.5,-0.3)--(-1.5,-0.3);
\draw[ very thick,blue](-0.5,0.3)--(-1.5,0.3);
\draw[ very thick,blue,->](0.5,0.3)--(1.5,0.3);
\draw[ very thick,blue,->](-2.5,0.3)--(-1.5,0.3);
\draw[ very thick,blue](1.5,0.3)--(3.5,0.3);
\draw[ very thick,blue](0.5,-0.3)--(1.5,-0.3);
\draw[ very thick,blue,->](3.5,-0.3)--(1.5,-0.3);
 \filldraw [black] (-0.5,0) circle (2pt);
  \filldraw [black] (0.5,0) circle (2pt);
  \filldraw [black] (1.5,0) circle (2pt);
   \filldraw [black] (2.5,0) circle (2pt);
 \draw[very thick,red](0,-2)--(0,2);
  \draw[very thick,red,->](0,0.5)--(0,1);
\draw[very thick,red,<-](0,-0.5)--(0,-1);
   \draw (0,-2.5) node{\small{$J=\frac{d-2}{2}+is$}};
    \draw (3,1.75) node{$J$};
    \draw (2.5,-0.75) node{$\Gamma$};
    \draw[thick](2.75,1.5)--(3.25,1.5);
    \draw[thick](2.75,1.5)--(2.75,1.9);
     \draw[very thick, blue] (-0.5,-0.3) arc (-90:90:0.3);
     \draw[very thick, blue] (-2.5,0.3) arc (90:270:0.3);
     \draw[very thick, blue] (0.5,0.3) arc (90:270:0.3);
\end{tikzpicture}
\hspace{3.0em}
\begin{tikzpicture}
\tikzset{decoration={snake,amplitude=.4mm,segment length=2mm,post length=0mm,pre length=0mm}}
\draw[thick,->](-2.5,-2)--(-2.5,2);
\draw[thick,->](-3.5,0)--(3.5,0);
 \filldraw [black] (-1.5,0) circle (2pt);
\filldraw [black] (-2.5,0) circle (2pt);
\draw[ very thick,blue](-2.5,-0.3)--(-1.5,-0.3);
\draw[ very thick,blue,->](-0.5,-0.3)--(-1.5,-0.3);
\draw[ very thick,blue](-0.5,0.3)--(-1.5,0.3);
\draw[ very thick,blue,->](0.5,0.3)--(1.5,0.3);
\draw[ very thick,blue,->](-2.5,0.3)--(-1.5,0.3);
\draw[ very thick,blue](1.5,0.3)--(3.5,0.3);
\draw[ very thick,blue](0.5,-0.3)--(1.5,-0.3);
\draw[ very thick,blue,->](3.5,-0.3)--(1.5,-0.3);
 \filldraw [black] (-0.5,0) circle (2pt);
  \filldraw [black] (0.5,0) circle (2pt);
  \filldraw [black] (1.5,0) circle (2pt);
   \filldraw [black] (2.5,0) circle (2pt);
 \draw[very thick,red](0.5,-2)--(0.5,-0.5);
  \draw[very thick,red,->](0.5,-2)--(0.5,-1);
 \draw[very thick,red,->](0.5,0.5)--(0.5,1);
  \draw[very thick,red](0.5,2)--(0.5,0.5);
  \draw[very thick, red] (0.5,0.5) arc (90:270:0.5);
   \draw (0,-2.5) node{\small{$J=\frac{d-2}{2}+is$}};
    \draw (3,1.75) node{$J$};
      \draw (2.5,-0.75) node{$\Gamma$};
    \draw[thick](2.75,1.5)--(3.25,1.5);
    \draw[thick](2.75,1.5)--(2.75,1.9);
     \draw[very thick, blue] (-0.5,-0.3) arc (-90:90:0.3);
     \draw[very thick, blue] (-2.5,0.3) arc (90:270:0.3);
     \draw[very thick, blue] (0.5,0.3) arc (90:270:0.3);
\end{tikzpicture}
\caption{The contour in the complex $J$ plane we meet during the Wick rotation is illustrated. The principal series for the Lorentzian spin is showed as the red line. The blue contour denoted as $\Gamma$ is the one we used to take the place of the sum over non-negative spin while the red line is the spin principal series. Wick rotation can be understood as the deformation of $\Gamma$ to the principal series. The figure on the left shows the configuration when the spacetime dimension $d$ is odd so that the principal series will not pass through any integer point on the real axes. The figure on the right shows the even dimension case in which we need to deform the principal series around $(\frac{d-2}{2},\;0)$ in order to make contours not meet with each other so that the Wick rotation is continuous.}
\label{f}
\end{figure}
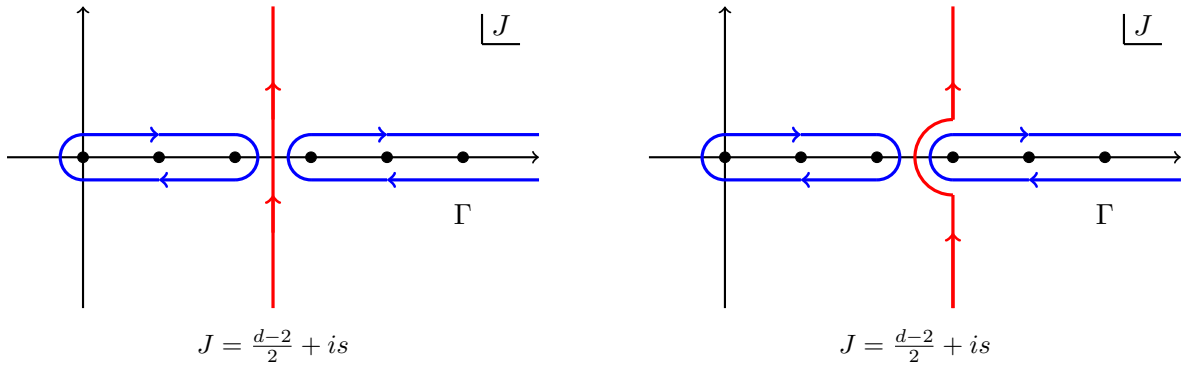We will also meet similar difficulty when doing rotation between two signatures. For example, in the study of partial wave decomposition of conformal four point functions \cite{Dobrev:1977qv,simmons2018spacetime,kravchuk2018light}, we need to  sum over all possible spin $J$ for every representations. This is represented as a sum over non-negative integers in Euclidean signature. In Lorentzian signature, in order to make such completeness still valid, we need to sum over the principal series, which now becomes a continuous integral. We see that such a gap between integer and continuous number arises again while this time it can be resolved by applying the complex analysis technics. \\\\   The key idea comes from the study of Sommerfeld-Waston transform \cite{eden2002analytic,gribov2003theory,cornalba2008eikonal}, which tells us that it is possible to rewrite the sum of discrete numbers into an integral along the proper contour in the complex $J$ plane. More precisely, we have
 \begin{equation}
     \sum_{J=0}^{\infty}\longrightarrow \frac{i}{2}\int_{\Gamma}\;\frac{dJ}{{\rm sin}(\pi J)},
 \end{equation}
in which the contour $\Gamma$ is shown in Fig.\ref{f}. At this stage, we are still in Euclidean signature and the next thing  we need to do is to deform the contour from $\Gamma$ to the Lorentzian principal series. During the deformation, various physical phenomenon will show up and it usually depends on the pole structure of the integrand. Specifically, we can write the integrand into the product of the coefficients $N_J(\Delta)$ and the basis $I^{E}_{\Delta, J}$ while the pole structure is hidden in the analytic extension of $N_J(\Delta)$, labelled by $N(\Delta,J)$. After figuring out the Lorentzian version of the basis $I_{\Delta,J}^L$, one may write down the formula in Lorentzian signature. We summarise the Wick rotation procedure into the formula
 \begin{equation}
    \sum_{J=0}^{\infty}\int_{\gamma}\; \frac{d\Delta}{2\pi i} \;N_{J}(\Delta)\; I^E_{\Delta,J}\longrightarrow \frac{i}{2}\int_{\gamma_J}\;\frac{dJ}{{\rm sin}(\pi J)}\int_{\gamma}\; \frac{d\Delta}{2\pi i} \;N(\Delta,J)\; I^L_{\Delta,J},\label{WR}
 \end{equation}
 which is the combination of following four steps\\\\ 
 i) Write the discrete sum $\sum$ in terms of the integral along $\Gamma$.\\\\
 ii) Do the space time rotation on the basis $I^E_{\Delta,J}$ in order to get the Lorentzian expression $I^L_{\Delta,J}$.\\\\
 iii) Analytically extend the coefficients $N_J(\Delta)$ to the complex plane to get $N(\Delta,J)$. \\\\
 iv) Rotate the contour $\Gamma$ to Lorentzian principal series $\gamma_J$ while the pole structure should be taken into consideration. \\\\
 The above method has already been applied to study the CFT partial wave decomposition \cite{Hartman:2015lfa,simmons2018spacetime,kravchuk2018light} and the poles are called Regge poles \cite{gribov2003theory,cornalba2008eikonal} in the context of scattering amplitudes. We will use it to deal with the conformal completeness relation in the next section.
\subsection{CFT Completeness Relation}
In conformal field theory, three-point correlation functions are elementary building blocks of the higher point functions and it is believed that, after imposing proper constraints, they will encode the whole information of a CFT following the idea of conformal bootstrap. Moreover, according to the harmonic analysis of the conformal symmetry group $SO(d+1,1)$ for a $d$ dimensional Euclidean CFT \cite{Dobrev:1977qv,Karateev:2018oml,Aharony:2020omh}, we can regard three-point functions as the basis of the fields and the orthogonality of the basis is given by the completeness relation
\begin{eqnarray}
    \delta(x_1,x_3)\delta(x_2,x_4)&=&\frac{1}{2}\sum_{J=0}^{\infty}\int_{\gamma}\frac{d\Delta}{2\pi i}\int d^dx_5\;{N_{J}}(\Delta)\label{come} \\\nonumber \\&\times&\langle O_{\Delta_0}(x_1)O_{\Delta_0}(x_2)O_{\Delta,J}^{\mu_1\cdots\mu_J}(x_5)\rangle\langle O_{\widetilde{\Delta}_0}(x_3)O_{\widetilde{\Delta}_0}(x_4)O^{\widetilde{\Delta},J}_{\mu_1\cdots\mu_J}(x_5)\rangle,\nonumber
\end{eqnarray}
in which $x_i^\mu$ for $0\le\mu \le d-1$ are spacetime coordinates and $\Delta$ is the scale dimension. Due to the harmonic analysis, the representation of $SO(d+1,1)$ will be unitary provided $\Delta$ lies on the principal series $\gamma=\frac{d}{2}+is$ and we denote the shadow transform of the operator $O_\Delta(x_i)$ as $O_{\widetilde{\Delta}}(x_i)$, in which $\widetilde{\Delta}=d-\Delta$. For simplicity, we will label them as $O_i$ and $\widetilde{O}_i$ in the later discussion. Moreover, given a CFT, we should note that the scale dimension usually has a lower bound named $\Delta_0$ and for free theory it takes the value $\frac{d-2}{2}$.\\ \\The completeness relation \eqref{come} can be used to expand local or bi-local fields in Euclidean signature while we need to do the Wick rotation on it to study the field theory in Lorentzian signature. In order to apply the Wick rotation skill, we first choose to integrate over $x_1, x_4$ on the two delta functions, making them a constant, and then focus on the integrand term on the RHS which is labelled as \footnote{More precisely, we should use $I^E_{\Delta,J}$.}
\begin{equation}
    I_{\Delta,J}:=\int d^dx_1d^dx_4d^dx_5\;\langle O_1O_2O_5\rangle\langle \widetilde{O}_3\widetilde{O}_4\widetilde{O}_5\rangle.
\end{equation}
The main task of the rest of this section is to find the expression of $I_{\Delta,J}$ in Lorentzian signature.\\ \\ There are five spacetime points in the product of two three-points functions and we use the conformal symmetry to fix three of them, i.e, we set $x_2=(0,\dots,0)$, $x_3=(1,0,\dots,0)$ and $x_5=(\Lambda,0,\dots,0)$\footnote{Usually we take $\Lambda\longrightarrow \infty$ but here we make $\Lambda$ a finite number.} and then obtain
\begin{equation}
    I_{\Delta,J}=\int d^dx_1d^dx_4\;\langle O_1O_2O_5\rangle\langle \widetilde{O}_3\widetilde{O}_4\widetilde{O}_5\rangle
\end{equation}
in which there are only two variables $x_1$ and $x_4$ left and they are explicitly shown in the integral. We should note that there will be an overall constant term arising from the volume of integral over $x_5$ when we choose the gauge but we do not need to consider that since it will go away once we ungauge the fixed points back to the integral in the end. To perform the Wick rotation, we introduce the normal Feynman continuation in which we take $x^0=(i+\epsilon)t$, $u=x^1-t$ and $v=x^1+t$. Therefore we obtain
\begin{equation}
    I_{\Delta,J}=-\frac{1}{4}\int dv_1du_1dv_4du_4d^{d-2}x_1d^{d-2}x_4\;\langle O_1O_2O_5\rangle\langle \widetilde{O}_3\widetilde{O}_4\widetilde{O}_5\rangle,
\end{equation}
and the singularities at the coincident points $x_1\thicksim x_2,x_5$, $x_4\sim x_3,x_5$ are given by
\begin{eqnarray}
u_{12}v_{12}+i\epsilon=0 \qquad u_{15}v_{15}+i\epsilon=0,\label{s1}\\\nonumber \\
u_{43}v_{43}+i\epsilon=0 \qquad u_{45}v_{45}+i\epsilon=0.\label{s4}
\end{eqnarray}
As for the integral, we can think about fixing $u_1$ and $u_4$ and then investigating the behavior of $I_{\Delta,J}$ on the $v_1$ and $v_4$ complex plane. Due to the introduction of the $i\epsilon$ expression, singularities will shift apart from the real axes and their positions are determined by the equation \eqref{s1} and \eqref{s4}. More explicitly, singularities of $v$ will shift to the upper half plane if $u$ is negative and vice versa. \\ \\Since we are interested in the nontrivial $I_{\Delta,J}$, the singularities $x_1\sim x_2$ and $x_1\sim x_5$ can not lie in the same half plane otherwise we can deform the integral contour along the real axes to the infinity and make $I_{\Delta,J}$ vanish. The same argument holds for the $x_4\thicksim x_3$ and $x_4\thicksim x_5$ singularities, which means that the sign of $u_{12}, u_{15}$ and $u_{43}, u_{45}$ should be different, i.e, it requires that
\begin{equation}
    u_2<u_1<u_5,\qquad u_3<u_4<u_5.
\end{equation}
\begin{figure}
\begin{tikzpicture}
\tikzset{decoration={snake,amplitude=.4mm,segment length=2mm,post length=0mm,pre length=0mm}}
\draw[thick,->](0,-2)--(0,2);
\draw[thick,->](-3.5,0)--(3.5,0);
 \draw[thick,decorate] (-3.5,-0.5) -- (0,-0.5);
  \draw[thick,decorate] (1,0.5) -- (3.5,0.5);
  \draw[thick,->](-3.5,-0.25)--(-2,-0.25);
  \draw[thick](-2,-0.25)--(0,-0.25);
  \draw[thick](-3.5,-0.75)--(-2,-0.75);
  \draw[thick,<-](-2,-0.75)--(0,-0.75);
   \filldraw [blue] (1,0.5) circle (2.5pt);
   \filldraw [blue] (0,-0.5) circle (2.5pt);
   \filldraw [blue] (2.5,0.5) circle (2.5pt);
   \draw[thick] (0,-0.75) arc (-90:90:0.25);
   \draw (-3,1.5) node{$v_1$};
   \draw (0.75,-1) node{\small{$1\thicksim2$}};
    \draw (0.75,1) node{\small{$1\thicksim5$}};
     \draw (3,1) node{\small{$1\thicksim4$}};
\end{tikzpicture}
\hspace{3.0em}
\begin{tikzpicture}
\tikzset{decoration={snake,amplitude=.4mm,segment length=2mm,post length=0mm,pre length=0mm}}
\draw[thick,->](-1,-2)--(-1,2);
\draw[thick,->](-3.5,0)--(3.5,0);
 \draw[thick,decorate] (-3.5,-0.5) -- (0,-0.5);
  \draw[thick,decorate] (1,0.5) -- (3.5,0.5);
  \draw[thick,->](3.5,0.75)--(2,0.75);
  \draw[thick](2,0.75)--(1,0.75);
  \draw[thick](3.5,0.25)--(2,0.25);
  \draw[thick,<-](2,0.25)--(1,0.25);
   \filldraw [blue] (1,0.5) circle (2.5pt);
   \filldraw [blue] (0,-0.5) circle (2.5pt);
   \filldraw [blue] (-1.5,-0.5) circle (2.5pt);
   \draw[thick] (1,0.75) arc (90:270:0.25);
   \draw (-3,1.5) node{$v_4$};
   \draw (0.75,-1) node{\small{$4\thicksim3$}};
    \draw (0.75,1.25) node{\small{$4\thicksim5$}};
     \draw (-2,-1) node{\small{$4\thicksim1$}};
\end{tikzpicture}
\caption{One possible choice of integral contour and branch cuts are shown in the $v_1$ and $v_4$ plane, in which blue points are coincident points in the correlation functions and wavy lines are branch cuts in the complex plane. In the $v_1$ plane the contour winding around $1\thicksim 2$ will generate the commutator $[O_1,O_2]$ while the contour around $4\thicksim 5$ in the $v_4$ plane will generate the commutator $[\widetilde{O}_5,\widetilde{O}_4]$. }
\label{sigu}
\end{figure}
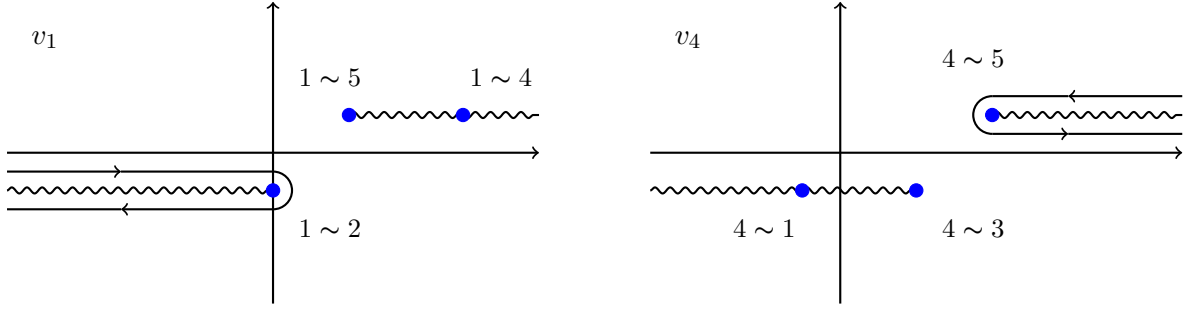
The next step is to determine the deformation of the integral contours of $v_1$ and $v_4$ and we illustrate one way of the deformation in Fig.\ref{sigu}, in which we let the $v_1$ integral go around $1\thicksim 2$ and the $v_4$ integral go around $4\thicksim 5$. This means that we restrict the integral in the region $v_1\leq 0, v_4\geq \Lambda$ and, at the same time, it induces the term 
\begin{equation}
    \langle[O_1,O_2]\;O_5\rangle\langle \widetilde{O}_3\;[\widetilde{O}_5,\widetilde{O}_4]\rangle,
\end{equation}
in which the winding of the contour around the branch cuts will produce a commutator $[,\;]$ and the order of the operator in the commutator is determined by the direction of the contour. After taking all kinds of deformation into consideration, we obtain
\begin{equation}
 ( \langle[O_1,O_2]\;O_5\rangle+\langle O_2\;[O_5,O_1]\rangle)\times (\langle \widetilde{O}_3\;[\widetilde{O}_5,\widetilde{O}_4]\rangle+\langle[\widetilde{O}_4,\widetilde{O}_3]\;\widetilde{O}_5\rangle),
\end{equation}
in which each term corresponds to a region in the $(v_1,v_4)$ plane and they together contribute to $I_{\Delta,J}$. After all, we should sum over all possible $\Delta$ and $J$ in $I_{\Delta,J}$ to get the completeness relation.  Now we should note that such problem has already been extensively discussed and resolved in \eqref{WR} and here we just write down the result in embedding space
\begin{eqnarray}
    &&\delta(P_1,P_3)\delta(P_2,P_4)= \int_{\gamma_J} \frac{idJ}{{\rm 2sin}(\pi J)}\int\frac{dP}{J!(\frac{d}{2}-1)_J} \int_{\gamma}\frac{d\Delta}{2\pi i}N(\Delta,J)\\\nonumber\\&\times& ( \langle[O_{\Delta_0}(P_1),O_{\Delta_0}(P_2)]\;O_{\Delta,J}(P,D_Z)\rangle+\langle O_{\Delta_0}(P_2)\;[O_{\Delta,J}(P,D_Z),O_{\Delta_0}(P_1)]\rangle)\nonumber\\ \nonumber\\ &\times& \langle O_{\widetilde{\Delta}_0}(P_3)\;[O_{\widetilde{\Delta},J}(P,Z),O_{\widetilde{\Delta}_0}(X_4)]\rangle+\langle[O_{\widetilde{\Delta}_0}(P_4),O_{\widetilde{\Delta}_0}(P_3)]\;O_{\widetilde{\Delta},J}(P,Z)\rangle)\nonumber+\cdots.
\end{eqnarray}
in which $P,Z$ are points in the embedding space corresponding to the boundary CFT, the overall constant $N(\Delta,J)$ will now become analytic function that depends on $\Delta,J$ and $\cdots$ represents the contribution from its poles. $D_Z$ is the operator used to contract the spin indexes associated with the factor $\frac{1}{J!(\frac{d}{2}-1)_J}$.
\subsection*{Physical Basis}
We have discussed that it is still unclear how to get physical spin from the principal series in Lorentzian signature while we found the connection between the Euclidean physical spin and the Lorentzian spin principal series via the Wick Rotation of the CFT completeness relation. Here we are going to directly rotate the physical Euclidean CFT to the physical Lorentzian CFT since spins are integers on both sides and we summarise all of these ideas in the following figure. 
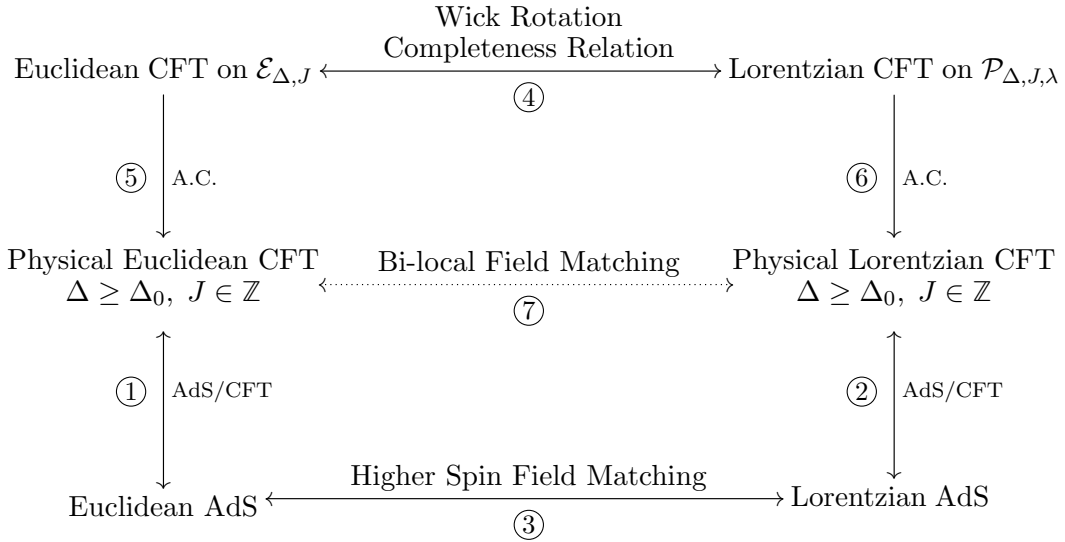
\begin{equation*}
\xymatrix{
{\rm Euclidean\; CFT\; on}\; \mathcal{E}_{\Delta,J}\ar[dd]^{\rm A.C.}_{\circled{5}}\ar@{<->}[rrr]^{\txt{Wick Rotation \\ Completeness Relation}}_{\txt{\circled{4}}} & {} &\qquad\qquad\qquad& {\rm Lorentzian\; CFT\; on\; } \mathcal{P}_{\Delta,J,\lambda}\ar[dd]^{\rm A.C.}_{\circled{6}} \\\\ *\txt{{}\\Physical Euclidean CFT \\$\Delta\geq \Delta_0,\; J\in \mathbb{Z}$\\{}}\ar@{<.>}[rrr]_{\circled{7}}^{\txt{Bi-local Field Matching}}\ar@{<->}[dd]^{\rm AdS/CFT}_{\circled{1}}&{} &&*\txt{{}\\Physical Lorentzian CFT \\$\Delta\geq \Delta_0,\; J\in \mathbb{Z}$\\{}} \ar@{<->}[dd]^{\rm AdS/CFT}_{\circled{2}}\\\\
\txt{Euclidean AdS} \ar@{<->}[rrr]^{\txt{Higher Spin Field Matching}}_{\circled{3}} & && {\rm Lorentzian\; AdS}\; }
\end{equation*}
The work about the analytic continuation of Euclidean CFT labelled by \circled{1} and Euclidean AdS/CFT map labelled by \circled{5} has already been discussed in \cite{Aharony:2020omh} while we have studied Wick Rotation of the completeness relation \circled{4} and high spin field matching \circled{3} in previous sections. In order to verify the AdS/CFT map in Lorentzian signature, denoted as \circled{2}, we need to find results in the physical Lorentzian CFT. Although the detail of analytic continuation of Lorentzian CFT is unclear, we find that it is possible to consider the direct correspondence between physical CFTs, labeled by \circled{7}.\\\\
\circled{7} can be regarded as the combination of \circled{4}, \circled{5} and \circled{6} and its net effect is just that we do the Wick rotation of space time while keep the spin physical. Now, we consider the behavior of spin in \circled{4} and \circled{6}, in \circled{4} we extend  physical Euclidean spin to Lorentzian principal series while in \circled{6} we need to extend spin on Lorentzian series to physical Lorentzian spin and in both cases the contour will move across poles on complex plane determined by $N(\Delta,J)$ but from opposite directions. Therefore we can see that the net contribution from the poles will be zero and then we obtain the physical CFT completeness relation written as  
\begin{eqnarray}
    &&\delta(P_1,P_3)\delta(P_2,P_4)=\sum _{J=0}^{\infty}\int \frac{dP}{J!(\frac{d}{2}-1)_J} 
    \int_{\gamma}\frac{d\Delta}{2\pi i}N(\Delta,J)\label{leq}\\\nonumber\\&\times& ( \langle[O_{\Delta_0}(P_1),O_{\Delta_0}(P_2)]\;O_{\Delta,J}(P,D_Z)\rangle+\langle O_{\Delta_0}(P_2)\;[O_{\Delta,J}(P,D_Z),O_{\Delta_0}(P_1)]\rangle)\nonumber\\ \nonumber\\ &\times&( \langle O_{\widetilde{\Delta}_0}(P_3)\;[O_{\widetilde{\Delta},J}(P,Z),O_{\widetilde{\Delta}_0}(P_4)]\rangle+\langle[O_{\widetilde{\Delta}_0}(P_4),O_{\widetilde{\Delta}_0}(P_3)]\;O_{\widetilde{\Delta},J}(P,Z)\rangle)\nonumber.
\end{eqnarray}
\subsection{AdS/CFT Map}
In this part, we will use the CFT completeness relation to expand bi-local fields $\Phi(P_1,P_2)$ on the boundary, together with expansion of higher spin fields on the bulk, then construct a map between them in Lorentzian signature so called Lorentzian AdS/CFT map . 
\subsubsection*{Bi-local Fields Matching}
Before going into the derivation of Lorentzian AdS/CFT map, we first discuss the match of bi-local fields during the Wick rotation, as the building block of our construction. First, following the convention from the previous section, we denote the bi-local fields in Euclidean and Lorentzian signature as $\Psi_E(P_1,P_2)$ and $\Psi_L(P_1,P_2)$, respectively. From the holography point of view, the match of bi-local fields is conceptually different from the match of bulk fields. But if we just consider the match of two theories with actions, the matching rule \eqref{mc} is universal. Therefore, consider the continuity of the field at the matching surface $\Sigma$, we immediately obtain the first matching condition, written as 
\begin{equation}
    \Psi_E(P_1,P_2)|_\Sigma=\Psi_L(P_1,P_2)|_\Sigma.
\end{equation}
We should note that such matching condition is universal and it does not depend on how we construct the bi-local fields while, for the second matching condition, one needs to consider the detail structure of the theory. First, we consider the $O(N)$ vector model in which the bi-local field is defined as
\begin{equation}
    \Psi(P_1,P_2):=\frac{1}{N}\sum_{I=1}^N\phi_I(P_1)\phi_I(P_2) ,
\end{equation}
where $\phi_I$ are real scalar fields. Since for each scalar $\phi_I$ one should impose the condition \eqref{m2}, we then obtain the second matching condition, written as 
\begin{equation}
    \partial_{\tau_a}\Psi(P_1,P_2)+i\partial_{t_a}\Psi_L(P_1,P_2)=0\qquad {\rm for } \qquad a=1,2.
\end{equation}
Then we consider the $U(N)$ vector model in which the bi-local field is given by 
\begin{equation}
    \Psi(P_1,P_2):=\frac{1}{N}\sum_{I=1}^N\phi^*_I(P_1)\phi_I(P_2) ,
\end{equation}
where $\phi_I$ become complex with the conjugate $\phi^*_I$ while the matching condition of the complex conjugate field $\phi^*_I$ is the complex conjugate of \eqref{m2}. Therefore, for the $U(N)$ vector model, the second matching condition now becomes \footnote{We should note that in the bi-local form $\phi$ and $\phi^*$ are not independent if one requires the bi-local field is hermitian, i.e., $\Phi^*(P_2,P_1)=\Phi(P_1,P_2)$. If we treat $\phi$ and $\phi^*$ independently, there will be no sign difference in the matching condition as derived in the Appendix \ref{csm}.  }
\begin{eqnarray}
    \partial_{\tau_1}\Psi(P_1,P_2)-i\partial_{t_1}\Psi_L(P_1,P_2)=0  \\\nonumber \\
     \partial_{\tau_2}\Psi(P_1,P_2)+i\partial_{t_2}\Psi_L(P_1,P_2)=0 ,
\end{eqnarray}
in which one can check that the matching conditions are hermitian in the sense that we treat 1,2 as the matrix indices. From the above discussion we can see that the explicit form of the second matching condition depends on the theory itself and there also exist physical models which do not require the second matching condition. For example, in SYK model the bi-local field is defined as 
\begin{equation}
        \Psi(P_1,P_2):=\frac{1}{N}\sum_{I=1}^N\langle\chi_I(P_1)\chi_I(P_2)\rangle ,
\end{equation}
in which $\chi_I$ are Majorana fermions. In this case, the requirement of the first matching condition is still necessary while there is no second matching condition since the on shell equation for $\chi_I$ is the first order differential equation. Moreover, one should impose proper reality and hermiticity condition on the spinor and action, respectively \cite{nicolai1978possible,van1996euclidean}.  
\subsubsection*{Derivation}
To make the Lorentzian results compatible with the Euclidean case, also for simplicity, we first define the Lorentzian three-point function basis as
\begin{eqnarray}
   \langle O_1O_2O_3\rangle_L:=  \langle[O_1,O_2]\;O_3\rangle+\langle O_2\;[O_3,O_1]\rangle,
\end{eqnarray}
\begin{equation}
    \langle \widetilde{O}_1\widetilde{O}_2\widetilde{O}_3\rangle_L:=  \langle \widetilde{O}_1\;[\widetilde{O}_3,\widetilde{O}_2]\rangle+\langle [\widetilde{O}_2,\widetilde{O}_1],\widetilde{O}_3]\rangle,
\end{equation}
in which we can see that Lorentzian three-point function basis are in fact combinations of three-point functions and its shadow counterpart is not just the shadow transform of each operator while we should also take the effect of commutator into consideration. But $\langle O_1O_2O_3\rangle_L$ and $\langle \widetilde{O}_1\widetilde{O}_2\widetilde{O}_3\rangle_L$ are still orthogonal in the sense of \eqref{leq}. Therefore, from the bi-local field expansion
\begin{eqnarray*}
    &&\Psi(P_1,P_2)_{E/L}=\sum_{J=0}^{\infty}\int_{\gamma_L}\frac{d\Delta}{2\pi i}\int \frac{dP}{J!(\frac{d}{2}-1)_J} \;\bar{C}_{\Delta,J}^{E/L}(P,D_Z) \langle O_{\Delta_0}(P_1),O_{\Delta_0}(P_2)O_{\Delta,J}(P,Z)\rangle_{E/L},
\end{eqnarray*}
 we can determine the coefficients $\bar{C}$ to be
\begin{eqnarray}
   && \bar{C}_{\Delta,J}^{E/L}(P,Z)=\frac{N_{\Delta,J}}{2}\int dP_1dP_2\; \Psi(P_1,P_2) \langle O_{\widetilde{\Delta}_0}(P_1),O_{\widetilde{\Delta}_0}(P_2)O_{\widetilde{\Delta},J}(P,Z)\rangle_{E/L}.
   \label{bdcoe}
\end{eqnarray}
As for the higher spin fields, we will focus on the study of transverse tensor fields and the transverse bulk completeness relation is given by
  \begin{equation*}
    \delta^{TT}(X_1,X_2)(W_{12})^J=\int_\gamma\frac{d\Delta}{2\pi i}\int \frac{dP}{J!(\frac{d}{2}-1)_J}\frac{N_{\Delta,J}}{\alpha_J}G_{\Delta,J}(X_1,P;W_1,D_Z)G_{\widetilde{\Delta},J}(X_2,P;W_2,Z),
\end{equation*}
which comes from the zero spin term of the full completeness relation and $\alpha_J$ are constants that depend on the spin. Therefore, given the off-shell tensor field expansion
\begin{equation}
    H(X,W)=\int_\gamma \frac{d\Delta}{2\pi i}\int \frac{dP}{J!(\frac{d}{2}-1)_J}C_{\Delta,J}(P,D_Z)G_{\Delta,J}(X,P;W,Z), \label{off-shell}
\end{equation}
we can use the bulk completeness relation to deduce the coefficient polynomial, which is given by
\begin{equation}
    C_{\Delta,J}(P,Z)=\frac{N_{\Delta,J}}{\alpha_J}\frac{1}{(\frac{d-1}{2})_JJ!}\int dX\; H(X,K_W)\;G_{\widetilde{\Delta},J}(X,P;W,Z).
    \label{bulkcoe}
\end{equation}
Now, we can construct a map between the bulk coefficients $C_{\Delta,J}$ and the CFT coefficients $\bar{C}_{\Delta,J}$, given by
\begin{eqnarray}
    C_{\Delta,J}^E(P,Z)=f^E_{\Delta,J}\;\bar{C}_{\Delta,J}^E(P,Z),\qquad  C_{\Delta,J}^L(P,Z)=f^L_{\Delta,J}\;\bar{C}_{\Delta,J}^L(P,Z), \label{map}
\end{eqnarray}
in which $f^E_{\Delta,J}$ and $f^L_{\Delta,J}$ are functions on the scale dimension and spin. Although one can propose the AdS/CFT map in Euclidean and Lorentzian signature separately, we will see that in fact $f^E_{\Delta,J}$ and $f^L_{\Delta,J}$ are not independent. According to the matching of bi-local field and the Wick rotation of the completeness relation, we know $\bar{C}^E_{\Delta,J}=\bar{C}^L_{\Delta,J}$ while we have $C_-^E=C_-^L$ from the study of higher spin field matching on the bulk. Taking these into consideration, we can conclude
\begin{equation}
    f_{\Delta,J}:=f^E_{\Delta,J}=f^L_{\Delta,J},
\end{equation}
which tells us that the AdS/CFT map should be invariant during the Wick rotation. Moreover, we can transfer the map between coefficients into the map between fields directly with the help of \eqref{bdcoe} and \eqref{bulkcoe}. We can deduce the CFT to AdS map to be
\begin{eqnarray}
    H(X,W)&=&\frac{1}{2}\int_{\gamma}\frac{d\Delta}{2\pi i}f_{\Delta,J}N_{\Delta,J}\int\frac{dP}{J!(\frac{d}{2}-1)_J}\int dP_1dP_2\;G^L_{\Delta,J}(X,P;W,D_Z)\nonumber\\\nonumber\\
    &\times& \langle O_{\widetilde{\Delta}_0}(P_1),O_{\widetilde{\Delta}_0}(P_2)O_{\widetilde{\Delta},J}(P,Z)\rangle_L\; \Psi(P_1,P_2),
    \end{eqnarray}
and the AdS to CFT map to be
\begin{eqnarray}
    \Psi(P_1,P_2)&=&\sum_{J=0}^{\infty}\int_\gamma\frac{d\Delta}{2\pi i}\frac{N_{\Delta,J}}{\alpha_J\;f_{\Delta,J}}\int\frac{dP}{(\frac{d-1}{2})_J(\frac{d}{2}-1)_JJ!^2}\nonumber\\ \nonumber\\
    &\times& \int dX\langle O_{\Delta_0}(P_1),O_{\Delta_0}(P_2)O_{\Delta,J}(P,Z)\rangle_L G^L_{\widetilde{\Delta},J}(X,P;K_W,Z)H(X,W),
\end{eqnarray}
therefore completes our derivation.\\ \\
Here, we should note that the AdS/CFT map provides us with a machinery to build connections between bi-local fields and higher spin fields and the fields could be off-shell and both take values in embedding space. In order to emphasis that the higher spin fields live on the bulk AdS and the bi-local fields live on the boundary CFT, we need to specify their physical regions. That is to say, the physical region for $H$ is the AdS surface $X^2=-1$ and the physical region for $\Psi$ is often taken to be light rays on the cone $X^2=0$. Moreover, for the on-shell AdS/CFT map, i.e, after solving equation of motion on the physical region, we will see that the on-shell AdS/CFT map is invertible in the large $N$ limit.

\section{Match Conditions for the Quadratic Action}\label{offshell}
As it was discussed in \cite{Aharony:2020omh}, the AdS/CFT map mainly works for the off-shell field while the matching condition introduced in section \ref{onshell matching} is applied to the on-shell field. In fact, on-shell fields belong to the subset of the off-shell fields. The expression \eqref{off-shell} of the off-shell fields is basically the expansion of an arbitrary function by the given basis $G_{\Delta,J}$ thus all the information of a physical systems is encoded in the coefficients $C_{\Delta,J}$. Suppose that the mass of the spin $J$ field is given by the scale dimension $\Delta_J$, although it is hard to solve $C_{\Delta,J}$ directly, we can deduce the coefficient will take the form 
\begin{equation}
    C_{\Delta,J}(P)\rightarrow \delta(\Delta-\Delta_J)C_{\Delta_J}(P), \label{pole}
\end{equation}
in which $C_{\Delta_J}(P)$ is the source of the spin $J$ field. Therefore, we can see that the mass spectra of the theory is determined by the pole structure of these coefficients $C_{\Delta,J}$, which will give us the AdS/CFT dictionary, together with the CFT coefficients $\bar{C}_{\Delta,J}$. So in order to complete the construction of Lorentzian AdS/CFT map, one needs to specify the matching conditions for the off-shell field.
\subsection{The Quadratic Action}
The match condition for a generic off-shell field is hard to study without specifying the equation of motion or the action for a given theory. In this article we will focus on the study of a special class of off-shell fields which govern the higher spin field theory dual to the vector model on the boundary and they are  described by the quadratic action shown in \cite{deMelloKoch:2018ivk,Aharony:2020omh}. For simplicity, we just consider the scalar action $S_E$ 
\begin{eqnarray}
S_E=\frac{1}{2}\int_{M_E} \;
\sqrt{G}(\partial^\mu\partial^\nu\Phi\;\partial_\mu\partial_\nu \Phi +M_1^2\;\partial_\mu\Phi\;\partial^\mu\Phi+M_2^2\;\partial_\nu\Phi\;\partial^\nu\Phi+M_1^2M_2^2\;\Phi^2) \label{qa}
\end{eqnarray}
in which $\Phi$ is the off-shell scalar field propagating in the manifold $M_E$ with AdS background described by the metric $G$ while $M_1$ and $M_2$ are mass of the particles. Here we should note that they are spin dependent parameters coming from the pole structure of $C_{\Delta,J}$ and fixed in our context since we are dealing with scalar fields setting $J=0$. Those off-shell scalar fields are not totally free while they should obey the equation of motion given by the action $S_E$, the variation is given by
\begin{eqnarray}
    \delta S_E &=& \int_{M_E}\sqrt{G}( \partial^\mu\partial^\nu\Phi\;\partial_\mu\partial_\nu \delta\Phi +M_1^2\;\partial_\mu\Phi\;\partial^\mu\delta\Phi+M_2^2\;\partial_\nu\Phi\;\partial^\nu\delta\Phi+M_1^2M_2^2\;\Phi\;\delta\Phi)\\ \nonumber \\
    &=&\int_{\partial {M_E}}\sqrt{G} \left(\left(2\partial_t\; \partial^i\partial_i \Phi+M_1^2\;\partial_t\Phi+M_2^2\;\partial_t\Phi\right)\delta\Phi+\partial^t\partial_t\Phi\;\delta\partial_t\Phi\right)\nonumber\\ \nonumber \\ &&+\int_{M_E} \left(\partial^\nu\partial^\mu \sqrt{G}\;\partial_\nu\partial_\mu\Phi-M_1^2\partial_\mu\sqrt{G}\;\partial^\mu\Phi-M_2^2\partial^\nu\sqrt{G}\;\partial_\nu\Phi+\sqrt{G}\;M_1^2M_2^2\Phi\right)\delta\Phi
\end{eqnarray}
in which the first term results from the boundary $\partial M_E$ and the second term gives rise to the equation of motion written as
\begin{eqnarray}
\frac{1}{\sqrt{G}}(\partial^\mu\partial^\nu\sqrt{G}\;\partial_\nu\partial_\mu\Phi-M_1^2\partial_\mu\sqrt{G}\;\partial^\mu\Phi-M_2^2\partial^\nu\sqrt{G}\;\partial_\nu\Phi)+M_1^2M_2^2\Phi&=&0.\label{quadratic}
\end{eqnarray}
Given that the AdS/CFT map is also valid in Lorentzian signature, the Lorentzian version of the action $S_L$ should exist and can be deduced to take the form of
\begin{eqnarray}
S_L=-\frac{1}{2}\int_{M_L} \;
\sqrt{-G}(\partial^\mu\partial^\nu\Phi\;\partial_\mu\partial_\nu \Phi +M_1^2\;\partial_\mu\Phi\;\partial^\mu\Phi+M_2^2\;\partial_\nu\Phi\;\partial^\nu\Phi+M_1^2M_2^2\;\Phi^2).
\end{eqnarray}
The matching condition of the total action $iS_L-S_E$ at the joint surface $\Sigma=\partial M_E=-\partial M_L$ is then deduced to be   
\begin{eqnarray*}
i\left(2\partial_t\;\partial^i\partial_i \Phi_L+M_1^2\;\partial_t\Phi_L+M_2^2\;\partial_t\Phi_L\right)\delta\Phi_L+\left(2\partial_\tau\;\partial^i\partial_i\Phi_E+M_1^2\;\partial_\tau\Phi_E+M_2^2\;\partial_\tau\Phi_E\right)\delta\Phi_E&=&0,\\ \\
i\partial^2_t\Phi_L\;\delta\partial_t\Phi_L-\partial^2_\tau\Phi_E\;\delta\partial_\tau\Phi_E&=&0,
\end{eqnarray*}
in which we have used the contraction relation $\partial_t^2=-\partial^t\partial_t$ and $\partial_\tau^2=\partial^\tau\partial_\tau$. Moreover, if we impose the condition that the charge $\partial^i\partial_i \Phi$ is conserved at the joint surface
\begin{equation}
    i\partial_t\partial^i\partial_i\Phi_L+\partial_\tau\partial^i\partial_i\Phi_E=0.\label{3 order}
\end{equation}
Then the matching conditions can be simplified to
\begin{eqnarray}
\Phi_E-\Phi_L=0,\label{0 order}\\\nonumber \\
i\partial_t\Phi_L+\partial_\tau\Phi_E=0,\label{1 order}\\ \nonumber\\
\partial^2_t\Phi_L+\partial^2_\tau\Phi_E=0,\label{2 order}
\end{eqnarray}
so called offshell matching conditions even though $\Phi_E$, $\Phi_L$ now satisfy the quadratic equation \eqref{quadratic}. Moreover, after rewriting the equation into the form of
\begin{equation}
    (\nabla^2_{AdS}-M^2_1)(\nabla^2_{AdS}-M_2^2)\Phi(X)=0
\end{equation}
with the $\nabla_{AdS}$ on AdS background. Then one can obtain the solution in terms of the bulk boundary propagator written as 
\begin{equation}
    \Phi(X)=\int dP \;\frac{1}{(X\cdot P)^{\Delta_1}}\;C_{\Delta_1}(P)+\int dP \frac{1}{(X\cdot P)^{\Delta_2}}\; C_{\Delta_2}(P),
\end{equation}
in which $\Delta_1$ and $\Delta_2$ are given by the relation
\begin{equation}
    M_1^2=\Delta_1(\Delta_1-d), \qquad M_2^2=\Delta_2(\Delta_2-d).
\end{equation}
$C_{\Delta_1}(P)$ and $C_{\Delta_2}(P)$ can be treated as coefficients or sources associated to the mode of scale dimension $\Delta_1,\Delta_2$. Therefore one can check the offshell match conditions \eqref{0 order} and \eqref{1 order} are solved by the matching of coefficients
\begin{equation}
    C_{\Delta_1}^E=C_{\Delta_1}^L,\qquad C_{\Delta_2}^E=C_{\Delta_2}^L \label{uni}
\end{equation}
and the condition \eqref{2 order} is then automatically preserved, which is shown in the appendix \ref{qm}. In fact, since we have seen that the offshell solution is the linear combination of two onshell fields, the offshell matching condition \eqref{3 order} can be rewritten in term of $\partial^3_t,\partial^3_t$ after using the onshell equation of motion for each propagator with mass $M_1,M_2$.\\ \\
Now we come back to the duality between the higher spin theory and the vector model. As it is known that the higher spin theory in the bulk is described by the Vasiliev equations and the corresponding action is still absent. Thus, with the help of AdS/CFT map, it is possible for us to reconstruct the bulk action from the action of boundary free vector model \cite{deMelloKoch:2018ivk,Aharony:2020omh}. This is done by first decomposing the boundary action into the coefficients $\bar{C}_{\Delta,J}$ then mapping it to bulk coefficients $C_{\Delta,J}$ via the AdS/CFT map \eqref{map} therefore rewriting the action in terms of the bulk fields. The spectra is determined by the pole structure of $C_{\Delta,J}$ as shown in \eqref{pole}. It turns out that the bulk action is quadratic consisting of higher spin fields and the scalar part is described by \eqref{qa} while the mass $M_1$ and $M_2$ are given by
\begin{equation}
    \Delta_1=d,\qquad \Delta_2=d-2.
\end{equation}
From the above procedure, we can see that one of the mode described by $C_{\Delta_1}$ is physical and the mode described by $C_{\Delta_2}$ with negative mass $M^2_2<0$ is unphysical based on the observation that it will contribute to the offshell field $\Phi$ in a negative way $C_{\Delta_2}<0$. Such unphysical modes are identified as ghost modes coming from the gauge fixing of the higher spin fields. Moreover, from the match condition \eqref{uni}, one can see that the physical and unphysical modes will behave independently during the Wick rotation.\\ \\
Furthermore, as pointed out in the work \cite{Skenderis:2009nt,Skenderis:2009kd} during the study of three-dimensional Einstein gravity, the higher-derivative terms in the action will introduce extra propagating degrees of freedom and they are identified as ghost modes. Those ghost modes make the theory unstable and violate the unitary condition. Such problem is rescued by considering the topologically massive gravity at the chiral point so that the left-moving sector will be gauge fixed and the extra degrees of freedom are then eliminated. Here for the higher spin fields, to make the theory physical, one can also choose to set the negative modes to zero in Euclidean signature by fixing the gauge of the higher spin fields , i.e $C_{\Delta_2}^E=0$. This naturally tells us that the ghost modes will not contribute to the external legs of the Feynman diagrams since the source is turned off in real time by checking the match condition \eqref{uni}. However, the ghost modes will contribute to the Feynman diagram at the loop level  while the detail goes beyond the study of classical match condition in this article \footnote{The matching condition developed in this article is the match for the classical field configuration and it tells us the vacuum is the Hartle-Hawking state. For the full description of the Hilbert space, one needs to study the match condition at quantum level.}. 
 \subsection{Analyticity}
 At first sight, it looks confusing since, starting from different actions, we eventually arrive at the same matching condition \eqref{uni} . As we will see, this becomes nature if one considers the behavior of the field and propagator over the whole complex time plane. The analytic property of the propagator $G_\Delta(X,P)$ implies that the condition \eqref{uni} could be universal for arbitrary higher order matching. \\\\
The complex time plane is described by the variable $z=t+i\tau$, $\bar{z}=t-i\tau$ and the function on the complex plane is denoted as 
\begin{equation}
    \Phi_C(z,\bar{z})=\Phi_C(t,\tau).
\end{equation}
\begin{figure}
\centering
\begin{tikzpicture}
\draw[thick,->] (0,4) --(0,2);
\draw[thick](0,2)--(0,0);
\draw[thick,->] (0,0)--(2,0);
\draw[thick](2,0)--(4,0);
\draw (-0.2,-0.2) node{o};
\draw (4,-0.3) node{$t$};
\draw (-0.3,4) node{$\tau$};
\draw(2,2)node{$\Phi_C(t,\tau$)};
\draw(-1,2) node{$\Phi_E(-\tau)$};
\draw(2,-0.5) node{$\Phi_L(t)$};
\end{tikzpicture}
\caption{The complexification of the field $\Phi_E$ and $\Phi_L$ into a function $\Phi_C$ with complex variables is depicted.}
\label{com func}
\end{figure}
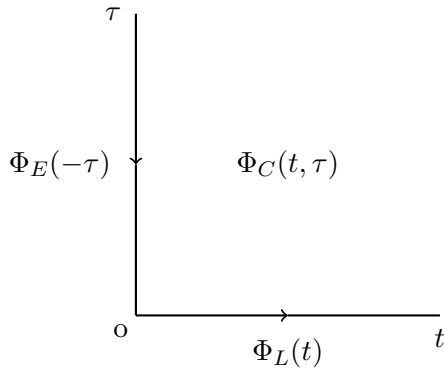\\
Although the complex time $t+i\tau$ will not make sense physically and we are just interested in the Euclidean field $\Phi_E(\tau)$ and the Lorentzian field $\Phi_L(t)$, we can still treat these two fields as living on the boundary of the complex field $\Phi_C(t,\tau)$. More precisely, as illustrated in Fig.\ref{com func} we can impose the condition
\begin{equation}
    \Phi_C(t,0)=\Phi_L(t),\qquad \Phi_C(0,\tau)=\Phi_E(-\tau),
\end{equation}
which specify the boundary value of $\Phi_C$ while the value in the interior is still unknown. But after taking the first order matching condition \eqref{1 order} into consideration
\begin{equation}
    \partial_t\Phi_L(t)-i\partial_\tau\Phi_E(\tau)=\partial_{\bar{z}}\Phi_C(t,\tau),
\end{equation}
we find that a sufficient condition for the complex field is the analyticity ,i.e, an analytic complex function will naturally induced a pair of field $\Phi_L$ and $\Phi_E$ which satisfy the first order matching. We can also see that the second order matching condition \eqref{2 order} can be written as $\partial_z\partial_{\bar{z}}\Phi_C=0$ thus now becomes trivial. Furthermore, using the relation
\begin{equation}
\partial_t=\frac{\partial_z+\partial_{\bar{z}}}{2} , \qquad \partial_t=i\frac{\partial_z-\partial_{\bar{z}}}{2},
\end{equation}
one can write the $\partial_t$, $\partial_\tau$ into the derivative $\partial_z$ and $\partial_{\bar{z}}$ and check that the third order matching condition \eqref{3 order} becomes trivial provided $\partial_{\bar{z}}\Phi_C=0$.\\ \\ Therefore, we can propose that the higher order matching condition are generated by the derivatives $\partial_z$, $\partial_{\bar{z}}$ acting on the complex function  and the matching condition for arbitrary higher order action is
\begin{equation}
     \label{ana}
    \partial_{\bar{z}}\Phi_C(z,\bar{z})=0.
    \end{equation}
This assumption is reasonable since as for the boundary quantum field theory, we have the analytic Wightman functions so we should expect to get analytic fields on the bulk via the help of holography principle. Although the Wightman functions are characterised by a series of axioms \cite{osterwalder1973axioms,osterwalder1975axioms} and here we start from the study of variation of the action $\delta S$. 

\section{Discussion}
In this paper, we have discussed various aspects of AdS/CFT map between the higher spin fields in the bulk and the bi-local fields on the boundary. We obtain the Lorentzian AdS/CFT map by doing the Wick rotation of higher spin fields, bi-local fields and the CFT completeness relation at the same time. Moreover, we should emphasize that the matching conditions are imposed at the linear and classic level and the complete matching for quantum gravity is still absent. The direct quantization of the two matching conditions we have determined in this article  depends on the matching of spectra and energy between two signatures. Furthermore, for a complete match of field theories at quantum level, more sophisticated conditions should be imposed and one can see the construction in the recent work \cite{kontsevich2021wick}.   \\ \\
Comparing to the Euclidean AdS/CFT map, though more complicated, the Lorentzian AdS/CFT map provides us with a powerful tool to study the real time duality, especially the study of black holes. Via the map, one can obtain the classic solution of the higher spin field in the bulk by dealing with the boundary CFT in the bi-local form, which is easier to work with since the boundary CFT is usually well understood. \\ \\
As for the boundary CFT, we have discussed the matching conditions for the $O(N)$ vector model, $U(N)$ vector model, and the SYK model in the context of vacuum-vacuum contour. It will be interesting to investigate how other contours, like thermal \cite{Christodoulou:2016nej,Pantelidou:2022ftm,Pantelidou:2023pjn} or out of time order contour, work in these theories and then possibly calculate the corresponding correlation functions.\\ \\
As we have discussed, both the mathematical structure and physical interpretation of the matching of the pole structure which show up on two sides of the AdS/CFT map is still unclear. It will be interesting to investigate this in some well studied models, for example, the duality between the JT gravity and the SYK model or the duality between the Higher spin on BTZ background and coset WZW models on the boundary. \\ \\
Moreover, the study of AdS/CFT map of finite $N$ may lead to a definition of quantum higher spin theory in the bulk since the theory of finite $N$ on the boundary is well defined. After getting the solution of the boundary theory of finite $N$, one can directly substitute it into AdS/CFT map the obtain a function on the bulk. This solution will carry the quantum information of the bulk theory, one can treat it as the solution of the bulk equation will some proper quantum corrected effective potential or use it to deduce the background then study the back reaction of the bulk geometry.\\ \\
As for fermion fields, the matching conditions will become more subtle since one needs to consider the representation of the symmetry group in different signatures and  fermions may transform in different ways, which makes it hard to impose the condition at the joint surface.  Moreover, given a Riemannian manifold $X$ with boundary $Y=\partial X$, one can classify the anomaly via studying the Dirac index by gluing a tube $Y\times \mathbb{R}_+$ to the boundary, which leads to the APS index theorem \cite{atiyah1975spectral,witten2016fermion}, while recently the theorem is formulated in the Lorentzian manifold \cite{bar2019index,bar2020local}. In our case, to prepare a Hartle-Hawking state, we need to study the Dirac operator on the Euclidean manifold with boundary but instead we should glue the Lorentzian manifold to the boundary $Y$. It is worthwhile to explore what the matching condition is in order to make the Dirac operator well defined.\\ \\




\section*{Acknowledgments}
I would like to thank my father Qinghe Hao and mother Xiulan Xu for providing funding for the tuition and accommodation fees when studying PhD at the University of Southampton. I also wish to thank Marika Taylor for the guidance throughout this project. I am grateful to Enrico Parisini and Kostas Skenderis for insightful discussions on ambient space and real time holography.

\appendix

\renewcommand{\theequation}{\Alph{section}.\arabic{equation}}

\setcounter{section}{0}

\section*{Appendix}
\setcounter{section}{0}

\section{Embedding space} \label{embedding}

\subsection{Bulk embedding space}

In this section we review the embedding of hyperbolic space and AdS space into flat space in one dimension higher, focussing in particular on the differences between Euclidean AdS (hyperbolic space) and Lorentzian AdS. \\ \\
We view $H^{d+1}$ as a spacelike surface in $R^{d+1,1}$. The coordinates of $R^{d+1,1}$ are denoted as $X = (X^+,X^-,X^{\mu})$ with $\mu = 0,\cdots (d-1)$, with the metric being:
\begin{equation}
ds^2 = - d X^- d X^+ + \delta_{\mu \nu} dX^{\mu} d X^{\nu}
\end{equation}
Then $H^{d+1}$ can be embedded as the spacelike surface $X^2 = -1$, which can be parameterized in Poincar\'{e} coordinates as
\begin{equation}
X = \frac{1}{z} \left ( 1, z^2 + y^2, y^{\mu} \right ). \label{poin}
\end{equation}
Note that as $z \rightarrow 0$ the embedding approaches $\frac{1}{z} P$ where $P$ is null. \\ \\
Lorentzian AdS$_{d+1}$ is embedded into $R^{d,2}$ as follows. The coordinates of $R^{d,2}$ are denoted as $X = (X^+,X^-,X^{\mu})$ with $\mu = 0,\cdots (d-1)$, with the metric being:
\begin{equation}
ds^2 = - d X^- d X^+ + \eta_{\mu \nu} dX^{\mu} d X^{\nu}
\end{equation}
AdS$_{d+1}$ can be embedded as the surface $X^2 = -1$ with signature (d,1), which can again be parameterized in Poincar\'{e} coordinates as \eqref{poin} but with the induced metric now being
\begin{equation}
ds^2 = \frac{1}{z^2} \left ( dz^2  + \eta_{\mu \nu} d y^{\mu} d y^{\nu} \right ).\label{metric}
\end{equation}
To distinguish different time directions it is convenient to use $t$ and $\tau$ to label the Lorentzian and Euclidean time directions; they share the same coordinate expression $t,\tau=y^0=X^0/X^+$. We will find it convenient to treat $X^0$ as a complex number with $t$ and $\tau$ the real and imaginary part, respectively.  In the embedding space the derivatives with respect to $t$ and $\tau$ can be expressed as
\begin{eqnarray}
    &&\frac{\partial}{\partial t}=\frac{\partial X^\mu}{\partial t}\frac{\partial}{\partial X^\mu}=-\frac{2t}{z}\frac{\partial}{\partial X^-}+\frac{1}{z}\frac{1}{\partial X^0}=-2X^0\frac{\partial}{\partial X^-}+X^+\frac{\partial}{\partial\label{dl} X^0},\\\nonumber \\
      && \frac{\partial}{\partial \tau}=\frac{\partial X^\mu}{\partial \tau}\frac{\partial}{\partial X^\mu}=\frac{2\tau}{z}\frac{\partial}{\partial X^-}+\frac{1}{z}\frac{1}{\partial X^0}=2X^0\frac{\partial}{\partial X^-}+X^+\frac{\partial}{\partial X^0}.\label{de}
\end{eqnarray}

\subsection{Conformal light cone}

In this section we summarise the embedding of flat space into the conformal light cone of a flat space in two dimensions higher, focussing on the differences between Euclidean and Lorentzian flat space. \\ \\
We denote the coordinates of the embedding space as $P = (P^+, P^-, P^{\mu})$ with $\mu = 0, \cdots (d-1)$. In the case of Euclidean $d$-dimensional flat space the metric on the embedding space $R^{d+1,1}$ is 
\begin{equation}
ds^2 = - d P^+ d P^- + \delta_{\mu \nu} dP^{\mu} dP^{\nu}
\end{equation}
The relation between the embedding coordinates and the flat space coordinates $x^{\mu}$ is determined by the conformal light cone conditions:
\begin{equation}
P^2 = 0 \qquad
P = \lambda P \label{conf}
\end{equation}
which are solved by $P = (1, x^2, x^{\mu})$ with $x^2 = x^{\mu} x_{\mu}$.  \\\\
For Lorentzian $d$-dimensional flat space the metric on the embedding space $R^{d,2}$ is 
\begin{equation}
ds^2 = - d P^+ d P^- + \eta_{\mu \nu} dP^{\mu} dP^{\nu}
\end{equation}
The relation between the embedding coordinates and the flat space coordinates $x^{\mu}$ is still determined by the conformal light cone conditions \eqref{conf}. 
\subsection{Solutions}
Here we briefly present the solutions that are used in section \ref{scalar} and one can see more detail in \cite{Skenderis:2008dg}. Given the real time action on $AdS_{d+1}$ background
\begin{equation}
    S=-\frac{1}{2}\int d^{d+1}x\sqrt{-G}(\partial_\mu \Phi\partial^\mu \Phi+m^2 \Phi^2)
\end{equation}
with the metric $G$ in \ref{metric}, we have the equation of motion
\begin{equation}
    z^{d+1}\partial_z(z^{-d+1}\partial_z\Phi)+z^2\square_0\Phi-m^2\Phi=0
\end{equation}
in which $\square_0$ is the Laplacian corresponding to the induced metric at $z=0$. In momentum space $q=(\omega, \vec{k})$, the solutions take the form 
\begin{equation}
    e^{-i\omega t+\vec{k}\cdot \vec{x}}z^{d/2}K_l(qz), \qquad    e^{-i\omega t+\vec{k}\cdot\vec{ x}}z^{d/2}I_l(qz), 
\end{equation}
in which $K_l$, $I_l$ are two types of Bessel functions. $\Delta=\frac{d}{2}+l$ is defined as $m^2=\Delta(\Delta-d)$. For spacelike momenta $q^2>0$ these modes are well behaved, while for timelike momenta, one need to specify how the contour of $\omega$ winds around the branch cuts of $q=\sqrt{q^2}$. Usually we use the $i\epsilon$ prescription
\begin{equation}
    q_{\epsilon}=\sqrt{-\omega^2+{\vec{k}}^2-i\epsilon}
\end{equation}
to specify that we are applying the Feynman contours $C$ over the complex $\omega$ plane. Moreover, the solutions behave like 
\begin{equation}
    z^{\frac{d}{2}}K_l\sim \frac{z^{d/2-l}}{q^l},\qquad z^{\frac{d}{2}}I_l\sim \frac{z^{\frac{d}{2}+l}}{q^{-l}}
\end{equation}
for $z\longrightarrow 0$ and
\begin{equation}
    z^{\frac{d}{2}}K_l\sim \sqrt{\frac{z^{d-1}}{q}}e^{-qz},\qquad z^{\frac{d}{2}}I_l\sim \sqrt{\frac{z^{d-1}}{q}}e^{qz}
\end{equation}
when $z\longrightarrow \infty$. According to their asymptotics behavior, $K_l$ are called source modes and contributes to the bulk-boundary propagator $X_L$ as 
\begin{equation}
    X_L(t,\vec{x},z)=\frac{1}{(2\pi)^d}\int_Cd\omega \int d\vec{k}e^{-i\omega t+i\vec{k}\cdot\vec{x}}\frac{2^{l+1}q^l_\epsilon}{\Gamma(l)}z^{\frac{d}{2}}K_l(q_\epsilon z),
\end{equation}
which leads to the familiar space time expression
\begin{equation}
    X_L=i\Gamma(l)\Gamma(l+\frac{d}{2})\pi^{-\frac{d}{2}}\frac{z^{l+\frac{d}{2}}}{(-t^2+\vec{x}^2+z^2+i\epsilon)^{l+\frac{d}{2}}}.
\end{equation}
As for $I_l$, they are called normalizable modes and we should note they will become infinite at $z=\infty$ when $q$ is space like. Therefore, we only consider the time like contribution 
\begin{equation}
    Y_L(t,\vec{x},z)=\frac{1}{(2\pi)^d}\int_Cd\omega \int d\vec{k}e^{-i\omega t+i\vec{k}\cdot\vec{x}}\;\theta(-q^2)b(\omega,\vec{k})\;z^{\frac{d}{2}}J_l(|q| z),
\end{equation}
in which $b(\omega,\vec{k})$ are undetermined coefficients and $J_l$ is defined as $I_l(-i|q|z)=e^{-i\pi l/2}J_l(|q|z)$. One can obtain the Euclidean version by doing the Wick rotation $t=-i\tau$. For source contribution we have
\begin{equation}
    X_E=i\Gamma(l)\Gamma(l+\frac{d}{2})\pi^{-\frac{d}{2}}\frac{z^{l+\frac{d}{2}}}{(\tau^2+\vec{x}^2+z^2+i\epsilon)^{l+\frac{d}{2}}}
\end{equation}
and the normalizable term is
\begin{equation}
    Y_E(t,\vec{x},z)=\frac{1}{(2\pi)^d}\int_Cd\omega \int d\vec{k}e^{-|\omega \tau|+i\vec{k}\cdot\vec{x}}\;\theta(-q^2)b(\omega,\vec{k})\;z^{\frac{d}{2}}J_l(|q| z),
\end{equation}
in which we have chosen the positive frequency when $\tau>0$. After lifting these to the embedding space, one should be able to recover the expressions shown at the beginning of section \ref{scalar} by making 
\begin{equation}
    G_\Delta=X+Y.
\end{equation}
\section{Solving BTZ}\label{btz}
We start from the Klein-Gordon equation for the scalar field $\Phi$, written as
\begin{equation}
    \nabla^2\Phi-m^2\Phi=0,
\end{equation}
in which the Laplacian operator $\nabla^2$ has the form of 
\begin{equation}
    \nabla^2\Phi=\frac{1}{\sqrt{-G}}\partial_\mu(\sqrt{-G}G^{\mu\nu}\partial_\nu \Phi).
\end{equation}
After using the BTZ black hole metric \eqref{BTZ metric}, we obtain the equation of motion in terms of the $(r,t,\phi)$ coordinates
\begin{equation}
    -\frac{1}{r^2-1}\partial_t^2\Phi+\frac{1}{r^2}\partial_\phi^2\Phi+(r^2-1)\partial_r^2\Phi+\frac{3r^2-1}{r}\partial_r\Phi-\Delta(\Delta-2)=0,
\end{equation}
in which we make $m^2=\Delta(\Delta-2)$. The modes of the above equation are given by 
\begin{equation}
    \psi=e^{i\omega t-ik\phi}f_{\Delta}(\omega,k,r),
\end{equation}
in which $\omega, k$ tell us how the modes propagate along the circle and $f_\Delta(\omega,k,r)$ is given by
\begin{equation}
    f_{\Delta}(\omega,k,r)=C_{\omega k\Delta}\left(1-\frac{1}{r^2}\right)^{-\frac{i\omega}{2}}r^{-\Delta}H(\frac{1}{r^2})
\end{equation}
where $C_{\omega k\Delta}$ are normalization constants while the  function $H(\frac{1}{r^2})$ is determined by Euler's hypergeometric differential equation
\begin{equation}
    z(1-z)H''+(\Delta-(\Delta+1-i\omega)z)H'-\frac{1}{4}(\Delta-i(\omega-k))(\Delta-i(\omega+k))H=0,
\end{equation}
where we have $z=\frac{1}{r^2}$ and the solutions are built from hypergeometric functions $F(a,b;c;z)$. Therefore in this case we have
\begin{equation}
    a=\frac{\Delta}{2}-\frac{i}{2}(\omega+k),\qquad b=\frac{\Delta}{2}-\frac{i}{2}(\omega-k),\qquad c=\Delta=l+1.
\end{equation}
Note that hypergeometric functions are locally expressed as power series and it converges when $|z|<1$ while the function over the whole complex $z$ plane can be obtained by the analytic continuation. Moreover, for physical systems, $\omega$ and $k$ are usually nonintegral thus the form of the solutions are mainly determined by the value of $\Delta$. Now, we discuss the solutions in two cases.\\ \\
i) \textit{None of the numbers $c$, $c-a-b$; $a-b$ is equal to an integer.}\\ \\
In this case, solutions can be expanded as combination of two independent power series at each singular point $z=0,1,\infty$.  Here we will only consider the behavior of the solution around the horizon $z=1$ and the infinity $z=0$. At $z=1$, solutions depend on the frequency $\omega$ and two independent solutions are given by
\begin{equation}
  H_{-\omega}(z)= F(\;\frac{\Delta}{2}-\frac{i}{2}(\omega+k),\;\frac{\Delta}{2}-\frac{i}{2}(\omega-k);\;1-i\omega;1-z)
\end{equation}
and 
\begin{equation}
  H_{\omega}(z)= (1-z)^{i\omega}F(\;\frac{\Delta}{2}+\frac{i}{2}(\omega-k),\;\frac{\Delta}{2}+\frac{i}{2}(\omega+k);\;1+i\omega;1-z).
\end{equation}
Taking these two into consideration at the same time, we can write the modes as
\begin{equation}
    \psi_{\pm}=e^{\pm i\omega t-ik\phi}f_{\Delta}(\pm\omega,k,r),
\end{equation}
in which
\begin{equation}
        f_{\Delta}(\pm\omega,k,r)=C_{\omega k\Delta}\left(1-\frac{1}{r^2}\right)^{-\frac{i\omega}{2}}r^{-\Delta}H_{\pm \omega}(\frac{1}{r^2})
\end{equation}
 At $z=0$, we also have two independent solutions while they now depend on the scale dimension $\Delta$ and we write them as 
\begin{equation}
    H_{\Delta^+}(z)=F(\;\frac{\Delta}{2}-\frac{i}{2}(\omega+k),\;\frac{\Delta}{2}-\frac{i}{2}(\omega-k);\;\Delta;z)
\end{equation}
and
\begin{equation}
    H_{\Delta^-}(z)=z^{1-\Delta}F(\;1-\frac{\Delta}{2}-\frac{i}{2}(\omega+k),\;1-\frac{\Delta}{2}-\frac{i}{2}(\omega-k);\;2-\Delta;z).
\end{equation}
Similar to the $z=1$ case, we can write the modes as 
\begin{equation}
    \psi_\pm=e^{i\omega t-ik\phi}f_{\Delta^{\pm}}(\omega,k,r),
\end{equation}
in which
\begin{equation}
    f_{\Delta^{\pm}}(\omega,k,r)=C_{\omega k\Delta}\left(1-\frac{1}{r^2}\right)^{-\frac{i\omega}{2}}r^{-\Delta}H_{\Delta^{\pm}}(\frac{1}{r^2}).
\end{equation}
ii) \textit{$c=\Delta=l+1$ is an integer for $l=1,2,3,\dots$}\\ \\
In this case, we will try to get the solutions of the equation from two ways. From one hand, we can apply the formula in \cite{abramowitz1988handbook} for integer $l$ directly and then obtain the fundamental system of the solution, given by
\begin{equation}
    H_{1(0)}(z)=F(a,b;l+1,z),
\end{equation}
and
\begin{eqnarray}
   && H_{2(0)}(z)=F(a,b;l+1;z)\;{\rm ln}z+\sum_{n=1}^m\frac{(a)_n(b)_n}{(1+l)_nn!}z^n(\psi(a+n)-\psi(a)+\psi(b+n)\\&&-\psi(b)-\psi(l+1+n)+\psi(l+1)-\psi(n+1)+\psi(1))-\sum_{n=1}^{\infty}\frac{(n-1)!(-m)_n}{(1-a)_n(1-b)_n} z^{-n}.
\end{eqnarray}
These are two independent solutions at $z=0$ and one can obtain the solution at $z=1,\infty$ through the analytic continuation.\\ \\ From the other hand, we can treat the integer case $\Delta=l+1$ as the limit of the general case $\Delta=l+1+\delta$ after taking $\delta \to 0$. This allows us to use the solutions for non-integer $\Delta$ thus their physical meanings are retained. To take the $\delta\to 0$ limit, we need to take care of the coefficients of the power series since they may have poles at integer $\Delta$. Taking the solution $H_{-\omega}$ for example. If we expand $H_{-\omega}$ around $z=0$ as power series, the term $\Gamma(k-l)$ will show up in the denominator of the coefficients of $z^k$ which leads to the poles when $k-l=1$. In fact, we can resolve those poles by choosing the hypergeometric function transformation $z\to 1-z$ for integer $\Delta$ then get
\begin{eqnarray}
    &&H_{-\omega}(z)=\frac{\Gamma(l)\Gamma(a+b-l)}{\Gamma(a)\Gamma(b)}z^{-l}\sum_{n=0}^{l-1}\frac{(a-l)_n(b-l)_n}{n!(1-m)_n}z^{n}-\frac{(-1)^l\Gamma(a+b-l)}{\Gamma(a-l)\Gamma(b-l)}\\\nonumber \\ &&\times\sum_{n=0}^{\infty}\frac{(a)_n(b)_n}{n!(m+n)!}z^{-n}\left( {\rm ln}z-\psi(n+1)-\psi(n+l+1)+\psi(a+n)+\psi(b+n)\right), 
\end{eqnarray}
from which we can see that the infinite term in the coefficients are transformed to the function ${\rm ln}z$. Given the above solution, we choose the normalization constants to be
\begin{equation}
    C_{\pm\omega k l}=\frac{\Gamma(\frac{1}{2}(l+1)+\frac{i}{2}(\pm\omega-k))\Gamma(\frac{1}{2}(l+1)+\frac{i}{2}(\pm\omega+k))}{\Gamma(l)\Gamma(1\pm i\omega)}
\end{equation}
so that the coefficient of the leading term $r^{l-1}$ in $\psi_\pm$ turns to be 1. Therefore, the  solution can be written as
\begin{equation}
    \psi=e^{\pm i\omega t -ik\phi}(r^{l-1}+\cdots +\alpha(\pm\omega,k,l)r^{-l-1}[{\rm ln}(r^2)+\beta(\pm\omega,k,l)]+\cdots),
\end{equation}
in which we just show the $r^{l-1}$ and $r^{-l-1}$ term and the coefficients $\alpha$, $\beta$ are given by
\begin{equation}
    \alpha(\pm \omega,k,l)=(-1)^l\frac{(\frac{i}{2}(\pm\omega+k)+\frac{1}{2}(1+l))_l(\frac{i}{2}(\pm\omega-k)+\frac{1}{2}(1+l))_l}{l!(l-1)!},
\end{equation}
\begin{equation}
    \beta(\pm \omega,k,l)=-\psi(\frac{i}{2}(\pm\omega+k)+\frac{1}{2}(1+l))-\psi(\frac{i}{2}(\pm\omega-k)+\frac{1}{2}(1+l))+\psi(1)+\psi(l+1).
\end{equation}
\section{Complex Scalar Matching} \label{csm}
In this section, we will derive the matching conditions for the complex scalar fields. We start from writing down the action for the complex scalar $\phi$. In Lorentzian signature we have
\begin{equation}
    S_L=\frac{1}{2}\int_{M_L} \sqrt{-G}(-\partial_\mu \phi_L^*\;\partial^\mu \phi_L-m^2\phi_L^*\phi_L)
\end{equation}
while for Euclidean signature the action is
\begin{equation}
    S_E=\frac{1}{2}\int_{M_E} \sqrt{G}(\partial_\mu\phi_E^*\;\partial^\mu \phi_E+m^2\phi_E^*\phi_E),
\end{equation}
in which we denote the Lorentzian spacetime and Euclidean spacetime as $M_L$ and $M_E$ joint at the codimension one surface $\Sigma$. As we have discussed, the continuation of the state implies the first matching condition $\phi_E=\phi_L$. Moreover, the stationarity of the total on-shell action with respect to $\phi$ and $\phi^*$ tells us 
\begin{equation}
   \frac{\delta}{\delta\phi}\; (\;iS_L-S_E)=\int_\Sigma \sqrt{K}(-i\partial_t\phi^*_L-\partial_t\phi^*_E)=0,
\end{equation}
\begin{equation}
   \frac{\delta}{\delta\phi^*}\; (\;iS_L-S_E)=\int_\Sigma \sqrt{K}(-i\partial_t\phi_L-\partial_t\phi_E)=0,
\end{equation}
in which $K$ is the intrinsic curvature induced on $\Sigma$. Therefore, we obtain the second matching condition written as 
\begin{equation}
    i\partial_t\phi_L+\partial_t\phi_E=0 \qquad {\rm and} \qquad   i\partial_t\phi^*_L+\partial_t\phi^*_E=0.
\end{equation}
\section{Quadratic Matching} \label{qm}
Here we present the detail of the verification of \eqref{2 order}. The Euclidean and Lorentzian propagator are given by 
\begin{equation}
    G_\Delta^E(X,P)=\frac{1}{(-2X\cdot P)_E^\Delta}\qquad G_\Delta^L(X,P)=\frac{1}{(-2X\cdot P)_L^\Delta}.
\end{equation}
Acting $\partial^2_t$ and $\partial_\tau^2$ on them, we obtain
\begin{equation}
    \partial_\tau^2 \frac{1}{(-2X\cdot P)_E^{\Delta}}=\frac{4\Delta(\Delta+1)}{(-2X\cdot P)_E^{{\Delta+2}}}(-X^0P^-+X^+P^0)^2-\frac{2\Delta}{(-2X\cdot P)^{\Delta+1}_E}X^+P^-
\end{equation}
and 
\begin{equation}
    \partial_t^2 \frac{1}{(-2X\cdot P)_L^{\Delta}}=\frac{4\Delta(\Delta+1)}{(-2X\cdot P)_L^{{\Delta+2}}}(X^0P^--X^+P^0)^2+\frac{2\Delta}{(-2X\cdot P)^{\Delta+1}_L}X^+P^-.
\end{equation}
Acting $\partial^3_t$ and $\partial^3_\tau$ on the propagator we have
\begin{equation}
    \partial_\tau^3 \frac{1}{(-2X\cdot P)_E^{\Delta}}=\frac{8\Delta(\Delta+1)(\Delta+2)}{(-2X\cdot P)_E^{{\Delta+2}}}(-X^0P^-+X^+P^0)^3-X^+P^-\frac{8\Delta(\Delta+1)}{(-2X\cdot P)^{\Delta+1}_E}(-X^0P^-+X^+P^0)
\end{equation}
and
\begin{equation}
    \partial_t^3 \frac{1}{(-2X\cdot P)_L^{\Delta}}=\frac{8\Delta(\Delta+1)(\Delta+2)}{(-2X\cdot P)_L^{{\Delta+2}}}(X^0P^--X^-P^0)^3+X^+P^-\frac{8\Delta(\Delta+1)}{(-2X\cdot P)^{\Delta+1}_L}(X^0P^--X^+P^0).
\end{equation}
One can directly check that \eqref{2 order} is true at the joint and surface $t=T$, $\tau=iT$, after taking the rotation of coordinates $P_0^E=iP_0^L$ into consideration.







\bibliographystyle{JHEP}
\bibliography{ref}

\end{document}